\documentclass[twocolumn]{aastex631}

\usepackage{graphicx}
\usepackage{dcolumn}
\usepackage{bm}
\usepackage{color}

\usepackage{newtxtext,newtxmath}
\usepackage{mathrsfs}
\usepackage[utf8]{inputenc}
\usepackage{graphicx}                           
\usepackage{amsmath}	
\usepackage[T1]{fontenc}
\usepackage{ae,aecompl}
\usepackage{ulem}
\begin{document}
\title[reconstruction of DM velocity]{AI-Powered Reconstruction of Dark Matter Velocity Fields from Redshift-Space Halo Distribution}

 \author{Xu Xiao}
 \affiliation{School of Physics and Astronomy, Sun Yat-Sen University, Zhuhai 519082, China}
 \author{Jiacheng Ding}
 \affiliation{School of Physics and Astronomy, Sun Yat-Sen University, Zhuhai 519082, China}
 \author{Xiaolin Luo}
 \author{Sun Ke Lan}
 \affiliation{School of Physics and Astronomy, Sun Yat-Sen University, Zhuhai 519082, China}
 \author{Liang Xiao}
 \affiliation{School of Physics and Astronomy, Sun Yat-Sen University, Zhuhai 519082, China}
 \author{Shuai Liu}
 \affiliation{School of Physics and Astronomy, Sun Yat-Sen University, Zhuhai 519082, China}
 \author{Xin Wang}
 \affiliation{School of Physics and Astronomy, Sun Yat-Sen University, Zhuhai 519082, China}
 \author{Le Zhang}
 \email{zhangle7@mail.sysu.edu.cn}
 \affiliation{School of Physics and Astronomy, Sun Yat-Sen University, Zhuhai 519082, China}
 \affiliation{Peng Cheng Laboratory, Shenzhen, Guangdong 518066, China}
 \affiliation{CSST Science Center for the Guangdong–Hong Kong–Macau Greater Bay Area, SYSU, Zhuhai 519082, China}
 \author{Xiao-Dong Li}
 \email{lixiaod25@mail.sysu.edu.cn}
 \affiliation{School of Physics and Astronomy, Sun Yat-Sen University, Zhuhai 519082, China}
 \affiliation{Peng Cheng Laboratory, Shenzhen, Guangdong 518066, China}
 \affiliation{CSST Science Center for the Guangdong–Hong Kong–Macau Greater Bay Area, SYSU, Zhuhai 519082, China}
\begin{abstract}

We propose a UNet-based deep learning model to reconstruct the real-space dark matter (DM) velocity field from the redshift-space distribution of sparse DM halos. Using various statistical measures, we show that the reconstructed velocity components--including velocity magnitude, momentum, and divergence--closely match the ground truth, achieving better than 10\% relative error and a correlation coefficient of 0.88. In the power spectrum comparison over $k \in [0.05, 0.3]~h/{\rm Mpc}$, the UNet reconstruction outperforms linear theory and agrees with the true field within $2\sigma$. The model also effectively corrects redshift-space distortions (RSD), yielding unbiased power spectrum  multipoles of DM fields within $2\sigma$. Notably, the UNet remains robust even with incomplete halo mass information. These results highlight the model’s broad applicability to cosmological analyses, including RSD, cosmic web studies, the kinetic Sunyaev-Zel'dovich effect, and BAO reconstruction.

\end{abstract}

\section{Introduction}\label{sec:intro}

The large-scale peculiar velocity is increasingly being recognized as a valuable tool in cosmology, as it is directly influenced by the gravitational pulling. This renders it invaluable for studying the dark universe~\citep{Peacock:2001gs,Linder:2003dr,Zhang:2007nk,Guzzo:2008ac,Jain:2007yk,Wang:2007ht,Reyes:2010tr,Li:2011sd,Clifton:2011jh,BOSS:2012xge,BOSS:2012grb,Weinberg:2013agg,Joyce:2014kja,Koyama:2015vza}, particularly its sensitivity to density fluctuations at horizon scales~\citep{Zhang:2010fa,Zhang:2015uta}, thereby providing insights into cosmic origin.

In a variety of circumstances, the measured velocity statistics may be given different weights. To illustrate, the kinetic Sunyaev-Zel'dovich (kSZ) effect is proportional to the gas momentum, which represents the gas density-weighted velocity. Conversely, the volume-weighted velocity power spectrum can be inferred from redshift-space distortions (RSD) by comparing the measured RSD power spectrum with theoretical models. In this approach, the theory of RSD includes the modelling of volume-weighted velocity statistics~\citep{Kaiser:1987qv,Scoccimarro:2004tg,Taruya:2010mx,Zhang:2012yt}.

In the context of cosmological studies, volume-weighted velocity statistics are considered to be a more appropriate approach than density-weighted ones. It is preferable to employ density-weighted statistics when utilizing the distribution function approach~\citep{2011JCAP...11..039S, 2012JCAP...11..014O} and the streaming model~\citep{1980lssu.book.....P, White:2014naa}. In contrast to density-weighted statistics, volume-weighted statistics significantly reduce the impact of uncertainties in galaxy density bias, though they do not eliminate it entirely. Nevertheless, the measurement of volume-weighted velocity statistics presents a number of challenges, both in observational studies and in numerical simulations. Velocities in regions containing galaxies can be determined; however, velocities in regions lacking galaxies (i.e., those with no simulation particles) typically remain present. This sampling artifact inevitably distorts the measurement of volume-weighted velocity statistics, with the impact increasing as the particle number density decreases. Additionally, there exists a correlation between the halos/galaxies spatial distribution and the velocity field being measured, due to an underlying correlation between the large-scale structure (LSS) and the latter. Consequently, the sampling of a volume-weighted velocity field is subject to bias, which results in a biased measurement of volume-weighted velocity statistics~\citep{Bernardeau:1995en,Bernardeau:1996hb,Schaap:2000se,Zheng:2013ora,zheng:2014vla,Zheng:2014ywa,Zhang:2014hra,Jennings:2014cqa}.

It has been observed that the velocity sampling artifact, which causes a suppression in the velocity power spectrum, becomes more significant as the sample number density, $\rho_h$, decreases. The effect is already discernible for $\rho_h \sim 1~(h/{\rm Mpc})^3$~\citep{Zheng:2013ora}. In the case of sparse samples, such as those observed in massive halos, the problem is considerably more pronounced. For a sample density of $\rho_h = 10^{-3}~(h/{\rm Mpc})^3$, the induced error in the velocity power spectrum reaches a value of approximately 10\%, even at scales as large as $k = 0.2~h / \mathrm{Mpc}$~\citep{Zheng:2014ywa}.

Several pioneering studies have been devoted to solving this long-standing problem. New velocity assignment methods have been developed, including the Voronoi tessellation method~\citep{Bernardeau:1995en}, the Delaunay tessellation method~\citep{Schaap:2000se}, the nearest-particle method~\citep{Zheng:2013ora,Koda:2015mca}, the Kriging method~\citep{Yu:2015gla,Yu:2016mzj}, and a hybrid approach to determine the volume-weighted halo velocity bias~\citep{Chen:2018ntg}. A theoretical model of the sampling artifact has been constructed~\citep{Zhang:2014hra}, validated in simulated DM velocity fields~\citep{Zheng:2014ywa}, and subsequently employed to correct the sampling artifact in the halo velocity field~\cite{zheng:2014vla}.

Sampling artifacts can significantly impact cosmological constraints when velocity power spectra are derived from sparse galaxy samples. This issue also affects RSD, a key cosmological probe used by Stage IV surveys such as DESI~\citep{levi2013desi}
EUCLID~\citep{laureijs2011euclid}, LSST~\citep{LSST},
WFIRST~\citep{WFIRST}, and CSST~\citep{2011SSPMA..41.1441Z}. These surveys would measure the volume-weighted velocity power spectrum through RSD with a statistical precision of approximately 1\% at  $0.1~h/{\rm Mpc}$. 

Both simulations and observations present challenges, prompting the question of {\it whether it is possible to accurately infer the density-weighted or volume-weighted velocity field directly from the spatial distribution of halos?} As this is a field-to-field mapping problem, it is non-trivial and complex. To date, there is no theoretical framework that can completely solve it. Given the large number of degrees of freedom in the field to be reconstructed, the traditional method is either invalid or inefficient for this purpose.

The recent advancements in machine learning algorithms, particularly those based on deep neural networks, present a significant opportunity to extract valuable insights from complex data. In more recent years, deep learning-based techniques have been applied with considerable success to almost all areas of cosmology and astrophysics~\citep{Mehta:2018dln,Jennings:2018eko,Carleo:2019ptp,Ntampaka:2019udw}, including weak gravitational lensing~\citep{Schmelzle:2017vwd,Gupta:2018eev,Springer:2018aak,Fluri:2019qtp,Jeffrey:2019fag,Merten:2018bgr,Peel:2018aei,Tewes:2018she}, the cosmic microwave background~\citep{Caldeira:2018ojb,Rodriguez:2018mjb,Perraudin:2018rbt,Munchmeyer:2019kng,Mishra:2019sep}, LSS for estimating cosmological parameters from the distribution of matter~\citep{Ravanbakhsh:2017bbi,Lucie-Smith:2018smo,Li2020...ML...2020SCPMA..63k0412P,2021JCAP...09..039L}, identifying DM halos, and reconstructing the initial conditions of the universe using machine learning~\citep{Modi:2018cfi,Berger:2018aey,Lucie-Smith:2019hdl,Ramanah:2019cbm}. In addition, this involves the mapping of coarse cosmology to the fine details~\citep{He:2018ggn,Li_2021}, extracting line intensity maps~\citep{Pfeffer:2019pca}, removing foregrounds in 21cm intensity mapping~\citep{Makinen:2020gvh}, augmenting N-body simulations with gas~\citep{Troster:2019mys}, mapping the 3D galaxy distribution in hydrodynamic simulations to its underlying DM distribution~\citep{Zhang:2019ryt}, modeling small-scale galaxy formation physics in large cosmological volumes~\citep{2021MNRAS.507.1021N}, reconstructing baryon acoustic oscillations~\citep{AIBAORecon...2020arXiv200210218M} and the initial linear-regime matter density field~\citep{2023MNRAS.520.6256S}, searching for gravitational waves~\citep{Dreissigacker:2019edy,Gebhar:2019ldz}, studying cosmic reionization~\citep{LaPlante:2018pst,Gillet:2018fgb,Hassan:2018bbm,Chardin:2019euc,Hassan:2019cal}, analyzing supernovae~\citep{Lochner:2016hbn,Moss:2018tug,Ishida:2018uqu,Li:2019ybe,Muthukrishna:2019wpf}, and conducting quasar studies~\citep{Monadi2023MachineLU,Jalan:2024aza,app15031024}.

The pioneering work by~\cite{Wu:2021jsy} demonstrated that a UNet architecture can accurately reconstruct the nonlinear velocity field of dark matter particles from the dark matter density field, achieving high precision down to scales of $2~h^{-1}{\rm Mpc}$. This result highlights the effectiveness of deep learning methods in capturing the complex dynamics of the cosmic velocity field at nonlinear scales. Subsequently,~\cite{Veena:2022now} employed an autoencoder-based UNet to reconstruct velocity fields from observed galaxy positions in dense redshift surveys.~\cite{Qin:2023dew} proposed a three-dimensional extension of UNet (called V-Net), for recovering velocity fields from redshift-space halo contrast maps.~\cite{Wang:2024ggp} applied a UNet model to SDSS DR7 data to reconstruct observational 3D peculiar velocity fields. In another approach,~\cite{Tanimura:2024igd} used Graph Neural Networks for velocity reconstruction of galaxy clusters. More recently,~\cite{Shi:2025zoz} employed a UNet network to recover the DM tidal field from mock catalogs of the DESI Bright Galaxy Survey. In addition, as demonstrated by~\cite{2021ApJ...913...76H,2023MNRAS.522.5291G}, the reconstruction of a density field or peculiar velocity field is possible from galaxy distributions.

Especially, \cite{2023MNRAS.522.4748W} have proposed a method for reconstructing the various peculiar velocity fields down to $k\lesssim 1.1$ $h/\rm Mpc$ from the redshift-space distribution of DM halos. However, reconstructing both the peculiar velocity and momentum fields of DM particles from the redshift-space distribution of sparse DM halos is more challenging, and has not yet been explored in previous studies.


In this study, we propose a UNet model dedicated to reconstructing the real-space density, velocity, and momentum fields from the redshift-space spatial distribution of DM halos (and subhalos). The latter two essentially correspond to the volume-weighted and density-weighted velocity fields. This study employs a simulation data set to investigate the effectiveness of a UNet-structured neural network in velocity reconstruction. The data set is introduced in Sect.~\ref{sec:method}, where the architecture of the neural network and the training procedure are also detailed. The results of the network are presented in Sect.~\ref{sec:result}, and finally, the conclusion and discussion are presented in Sect.~\ref{sec:con}. 

\section{Methods}\label{sec:method}
\subsection{Dataset}
In order to train and validate our deep learning model, we employed the CosmicGrowth simulations~\citep{cosmicgrowth} as our training and test datasets. The simulations are high-resolution N-body simulations with 8.6 or 29 billion particles, based on the flat $\Lambda$CDM model. Cosmological parameters are derived from the WMAP~\citep{WMAP2011,WMAP2013} or Planck results~\citep{Planck:2013pxb}. In addition to standard $\Lambda$CDM runs, the suite includes simulations with two alternative CDM models and a range of initial power spectra, with the scalar spectral index varying from $n\in[ -2.0,0.0]$. Specifically, in this study, we adopted a WMAP-based simulation with the cosmological parameters of $\{\Omega_c, \Omega_b, h, n_s, \sigma_8\} = \{0.2235, 0.0445, 0.71, 0.968, 0.83\}$, consistent with those from Planck~\citep{Planck:2013pxb,Planck:2018vyg} within the $2\sigma$ level. Therefore, we do not expect this choice to have a significant impact on the performance of velocity reconstruction.

This simulation was performed in a cubic volume with a side length of 1.2~$h^{-1}{\rm Gpc}$, containing $2048^3$ DM particles, each with a mass resolution of $M_{\rm DM} = 3.8 \times 10^{9}~h^{-1}M_{\odot}$. DM halos in each snapshot of the CosmicGrowth simulations were identified using the Friends-of-Friends (FoF) algorithm, and the corresponding merger trees and subhalo catalogs were constructed with the Hierarchical Branch Tracing (HBT) algorithm~\citep{HBT2012}.

Our work is conducted in preparation for upcoming CSST-like Stage-IV slitless spectroscopic galaxy surveys, which primarily target galaxies at redshifts $z < 1$~\citep{Gong_2019}. As a representative case, we constructed the halo and subhalo catalog in redshift space at $z = 0.59$. The input data were required to have a number density of $0.003~h^3{\rm Mpc}^{-3}$, which corresponds to a minimum halo mass of $M_{\rm min} \simeq 1.4 \times 10^{12}~h^{-1}M_\odot$. This threshold is consistent with current spectroscopic galaxy surveys. 

The relationship between the redshift space position, denoted by $\bm{s}$, and the real space position, denoted by $\bm{r}$, after accounting for the RSD effect is given by 
\begin{equation}\label{eq:rsd}
\bm{s}=\bm{r}+\frac{\bm{v} \cdot \hat{z}}{a H(a)} \hat{z}\,,
\end{equation}
where $\bm{v}$ denotes the peculiar velocity, $a$ denotes the scale factor, $H(a)$ denotes the Hubble parameter, and $\hat{z}$ represents the unit vector along the line of sight (LoS). The density and velocity fields were computed based on the catalogue samples, with the haloes assigned to a $512^3$ mesh using the CIC (Cloud-in-Cell) scheme, with a cell resolution of $2.35 h^{-1}~{\rm Mpc}^3$.

\subsection{Training process}
\begin{figure*}[htpb]
	\centering
	\includegraphics[scale=0.65]{"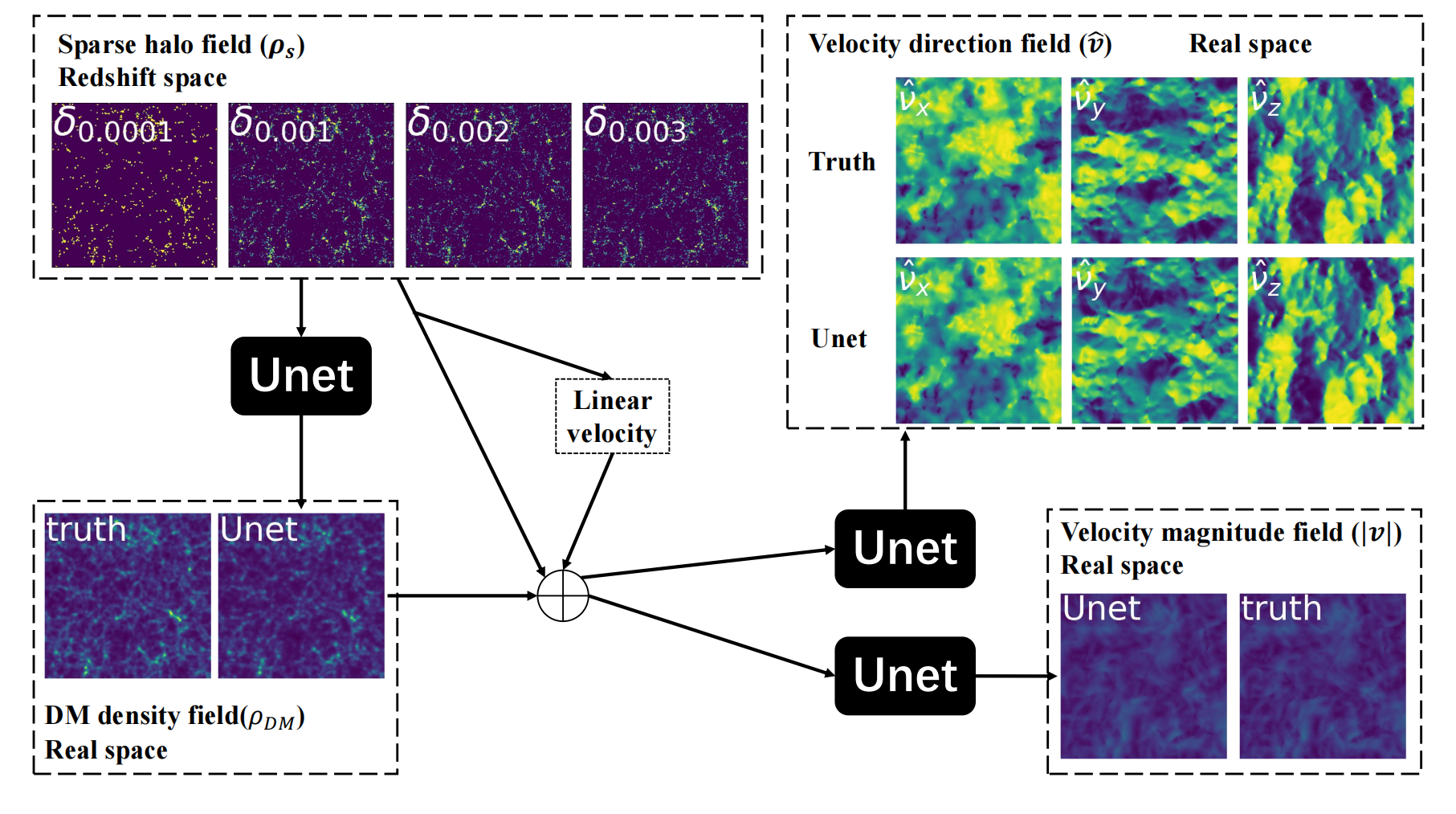"}
	\caption{Training scheme for reconstructing the DM velocity field in real space. The reconstruction process involves two major steps. The first step is to reconstruct the DM density field, $\rho_{\rm DM}$, from the sparse DM halo density field, $\rho_s$, in redshift space. To achieve this, we trained a UNet model to predict $\rho_{\rm DM}$. Subsequently, we trained another UNet model that takes as input the concatenated data from the linear velocity prediction. This model is then used to reconstruct the direction and magnitude fields of velocity. This network structure is also applicable when the final output is the real-space momentum field.}
	\label{fig:model_processing}
	\end{figure*}

The objective is to construct a mapping based on the UNet method, which will take as its input a sparse halo number density field in redshift space (to provide an accurate representation of real observations) and output the real-space fields, including the DM density field and DM velocity/momentum fields. 

For clarification, let us first define two density fields that will be used during the UNet reconstruction process: 1) {\it the redshift-space sparse halo number density field}, $\rho_s$; 2) {\it the real-space DM density field}, $\rho_{\rm DM}$.

Moreover, the  real-space comoving velocity field is denoted by $\bm{v}$, with its magnitude represented by $|\bm{v}|$ and its direction by $\hat{v}$. The momentum field in real space is defined as the number-weighted velocity, given by $\bm{m} = (1 + \delta_{\rm DM}) \cdot \bm{v}$, where $\delta_{\rm DM} \equiv \rho_{\rm DM} / \bar{\rho}_{\rm DM} - 1$ is the DM density field relative to its mean. 

The reconstruction process is divided into two principal steps, as illustrated in Fig.~\ref{fig:model_processing}, which provides a detailed illustration of the UNet network in Fig.~\ref{fig:model_structure}. For the illustrative purpose, the velocity will be taken as an example. The same methodology is employed for the processing of the momentum field. 

In what follows, we will provide a comprehensive description of this process. 



1) The mapping from $\rho_s(\bm{x})$ to $\rho_{\rm DM}(\bm{x})$.  The initial step is the reconstruction of the distribution of DM in real space, $\rho_{\rm DM}$, using the sparse halo number density field in redshift space, $\rho_s$. Each channel of the input contains halos within a specific mass range. The DM halos (and subhalos) are catalogued in descending order of mass and divided into four mass intervals, with boundaries defined by the logarithmic mass ratio as follows: $\log_{10}(M/M_\odot) \in [15.01,13.30, 12.56, 12.31, 12.17]$, corresponding to four number densities of halos $\rho_h\in [0.0001, 0.001, 0.002, 0.003]~(h/{\rm Mpc})^3$. Note that the mass information may be estimated to a reasonable degree of accuracy using empirical formulas, which allow for the inference of the masses of these objects from the observed apparent magnitudes of galaxies. The target data is the three-dimensional number density field of DM in real space. Our UNet model is constructed for the purpose of reconstructing the real-space $\rho_{\rm DM}$. 

Two mass-weighting schemes are adopted for comparison:
i) ``with $M_{\rm halo}$ weighting'': in this scheme, DM halos are divided into the four intervals based on mass, and each interval is assigned a weight proportional to its mass.

ii) ``without $M_{\rm halo}$ weighting'': in this scheme, DM halos are also divided into the four intervals based on mass, but no weights are assigned to these intervals.

The first scheme assumes an ideal scenario where we have complete knowledge of the mass information of DM halos. In contrast, the second one better reflects current observational capabilities, allowing only rough estimates of the mass distribution of galaxies, and thus corresponds more closely with actual observations. To assess the impact of mass information accuracy, results from both schemes will be presented simultaneously.

2) The mapping from a combination of $\rho_s(\bm{x})$ and $\rho_{\rm DM}(\bm{x})$ to $\bm{v}(\bm{x})$ or $\bm{m}(\bm{x})$ is achieved using two UNet models. These models are constructed to reconstruct the direction and magnitude of the DM velocity and momentum fields in real space. The UNet models utilize $\rho_s$, $\rho_{\rm DM}$, and the linear velocity in redshift space, $\bm{v}_{\rm lin}$, derived from linear perturbation prediction as inputs.

\subsection{Neural network model}

\begin{figure*}[htpb]
	\centering
	\begin{tabular}{cc}
		\includegraphics[height=0.28\textwidth]{"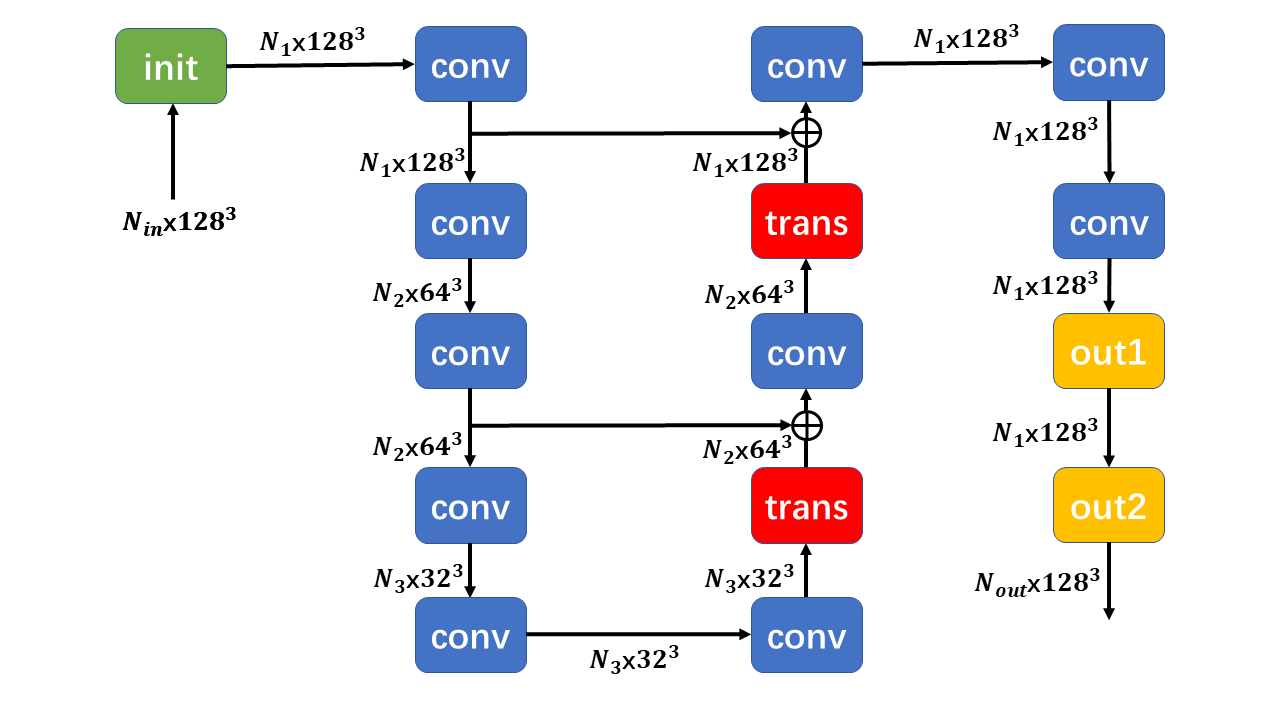"} &
		\includegraphics[height=0.28\textwidth]{"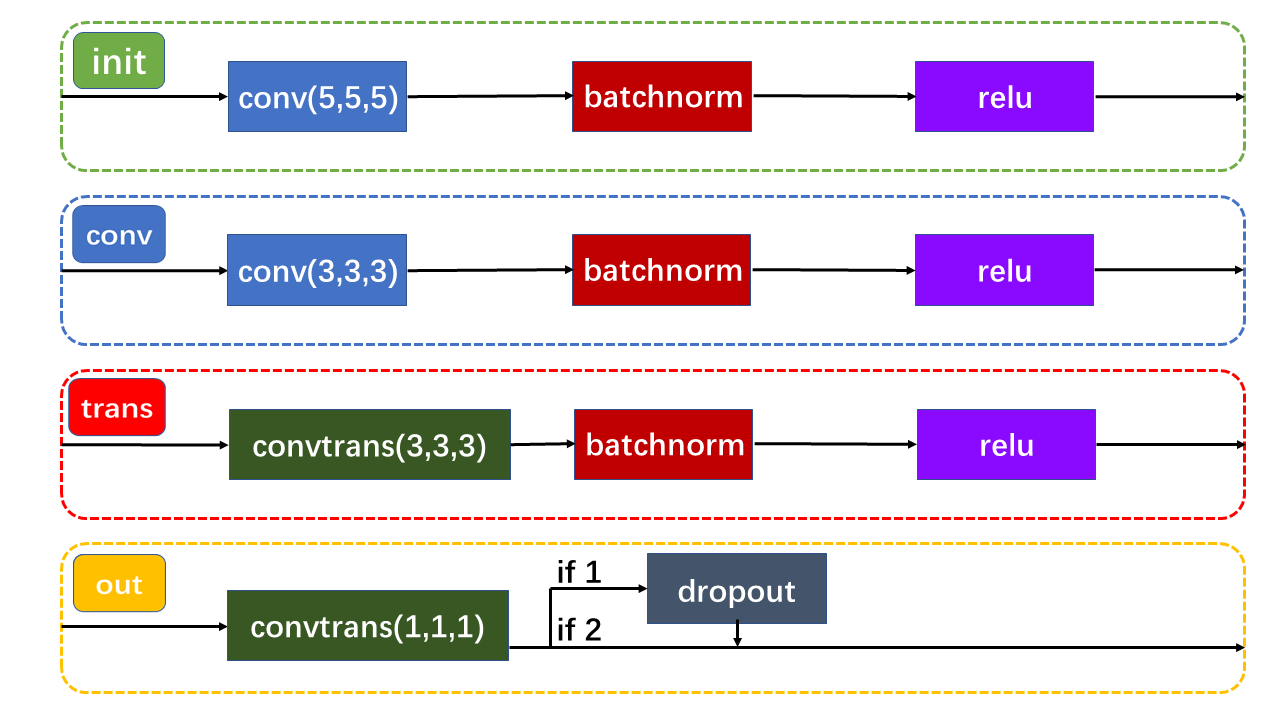"} \\
	\end{tabular}
\caption{UNet neural network architecture employed for the reconstruction of the DM density and velocity/momentum fields. A $128^{3}$ box with a side length of 300 ${\rm Mpc}/h$ is employed in the UNet architecture as the input. This architecture is comprised of a three-block structure: a up-convolution block, a down-convolution block, and a final convolution block. The output fields have dimensions of $128^3$, with a volume of $300^3~({\rm Mpc}/h)^3$. The left panel provides detailed information on the components included in the three blocks of the UNet, while the right panel offers detailed descriptions of the ``init'', ``conv'', ``trans'', and ``output'' layers. In the final convolution block, a dropout layer is employed between the convolutional transformation and batch normalization layers to enhance the performance of UNet and prevent overfitting. The values of $N_1, N_2, N_3$, and the dropout percentage are automatically provided by a machine learning tool called Optuna~\citep{optuna2019}.}
    \label{fig:model_structure}
	\end{figure*}

In light of the neural network model proposed by~\cite{Wu:2021jsy}, we employed the UNet architecture for the construction of our model. The architecture of our neural network and its components are depicted in Fig.~\ref{fig:model_structure}. The input is data blocks with an arbitrary number of channels, with each channel corresponding to a specific mass range of $\rho_s$. 

Given that the velocity field is decomposed into the  velocity magnitude and direction, we constructed two structurally similar neural networks to handle them separately. The networks conclude with an output layer comprising 1+3 channels. The three channels correspond to the components of the velocity direction, while the fourth channel corresponds to the velocity magnitude. The final step in the reconstruction process is the combination of the results from all output channels, which ultimately yields the complete reconstruction of the three-dimensional velocity field.

 The notations in Fig.~\ref{fig:model_structure}, ``${n_{\rm in}}$'' and ``${n_{\rm out}}$'', represent the number of input and output channels, respectively. These values may vary across different reconstruction tasks, as shown in Tab.~\ref{tab:channel}.

\begin{table}[!h]
 \centering
 \caption{Input and output channel numbers based on different reconstructed fields.}
\begin{tabular}{c|c|c}
	\hline
            fields & ${n_{\rm in}}$  & ${\rm n_{\rm out}}$ \\
	\hline
		DM density ($\rho_{\rm DM}$)& 4 & 1 \\
        magnitude of  DM velocity ($|\bm{v}|$) & 8 & 1 \\
        direction of DM velocity ($\hat{v}$) & 8 & 3 \\
        magnitude of  DM momentum ($|\bm{m}|$) & 8 & 1 \\
        direction of DM momentum ($\hat{m}$) & 8 & 3 \\
	\hline
 \end{tabular}\label{tab:channel}
\end{table}
    
The color blocks represent the various operations performed by the neural network, with arrow lines connecting the input and output. The dimensions of the input, intermediate, and output fields (channels $\times$ spatial pixels) are defined, along with the size and number of labeled 3D convolutional kernels (``conv'').  

Given the constraints of GPU memory, training time, and model size, we divided a large box with a physical scale of 1200 ${\rm Mpc}/h$ into $4^3$ small boxes with a physical scale of 300 ${\rm Mpc}/h$, each containing $128^3$ grid points in the CIC interpolation scheme. 
 The 300 ${\rm Mpc}/h$ sub-boxes are sufficiently large to capture all relevant nonlinear velocity features at scales $\lesssim 40~{\rm Mpc}/h$ and linear-theory-predicted features at scales $\gtrsim 100~{\rm Mpc}/h$, while remaining compact enough to satisfy the computational constraints of model training.

We utilized a total of 64 small boxes during the training and validation phases--32 boxes for training and 32 for validation. Each box had dimensions of $128^3$ pixels, corresponding to a side length of $300~{\rm Mpc}/h$. For testing, we used boxes with dimensions of $256^3$ pixels, corresponding to a side length of $600~{\rm Mpc}/h$.

Note that, due to GPU memory limitations, we trained on smaller $128^3$ pixel boxes, as using $256^3$ boxes with typical batch sizes would exceed our GPU capacity. A comparison with~\cite{Wang:2023hgm} shows that our method, despite using smaller training boxes, achieves comparable reconstruction accuracy, indicating no significant loss in performance. Using the neural network on large boxes helps us eliminate less accurate outputs at the edges, which can be a significant issue when applying the UNet model to smaller boxes. For training, the $k_{\rm min}$ for small boxes is 0.026, while the $k_{\rm min}$ for validation with large boxes is 0.013.


Some points require further clarification, which will be detailed below.
\begin{enumerate}
\item[1)]  
To modify the field size, the $\emph{stride}=2$ parameter can be employed, which allows for a reduction or increase in the field size by a factor of two.

\item[2)] 
The ``init'' 3D convolutional layer enables the network to rapidly learn large-scale information due to its sufficiently large receptive field.

\item[3)]
The convolutional layer, which is situated at the output of the network, serves to increase the network's complexity. In order to prevent overfitting, a dropout layer is incorporated at the end of the convolutional layer. This layer also alters the number of channels.

\item[4)]
The incorporation of batch normalization (BN) layers into the neural network facilitates the acceleration of training convergence and the prevention of overfitting. Additionally, the implementation of rectified linear units (ReLU) as activation layers subsequent to convolutional layers enhances the network's nonlinearity. 


\item[5)]

Additionally, due to the constraints of the limited dimensions of the training box, the model struggles to learn large-scale velocity modes. To mitigate potential bias from the small box dataset, the velocity field predicted by traditional linear reconstruction is used as one of the UNet inputs to reconstruct the velocity or momentum field. The linear velocity and momentum predictions in redshift space are determined through
\begin{equation}\label{eq:linear}
\bm{v}_{\rm lin}(\bm{k})=a f H \frac{i \bm{k}}{k^2} \frac{\delta_s(\bm{k})}{b}\,,
\end{equation}
and 
\begin{equation}\label{eq:linear_momentum}
\bm{m}_{\rm lin}=(1+\frac{\delta_s}{b})\bm{v}_{\rm lin}\,.
\end{equation}
Here $\delta_s$ is divided by the linear bias $b$, where $b=$ 1.85 for ``without $M_{\rm halo}$ weighting'' scheme and 2.57 for ``with $M_{\rm halo}$ weighting'' scheme and is determined by the ratio of the integrated DM power spectrum to the integrated halo power spectrum over the relevant $k$-range of interest, i.e., 
\begin{equation}
b^2=  \frac{\int_{k<0.3}\frac{ P_{\rm halo}(k) }{P_{\rm DM}(k)}dk}{\int_{k<0.3}dk}\,.
\end{equation}


\item[6)]
The Tree Parzen Estimator (TPE) process~\citep{NIPS2011_86e8f7ab}, based on Bayesian optimization, is employed to optimize hyperparameters, with the Python package Optuna\footnote{https://optuna.readthedocs.io} utilized to maximize the model accuracy~\citep{DBLP:journals/corr/abs-1907-10902}. The hyperparameters that are optimized include:  1) the learning rate; 2) dropout percentage; 3) the number of training data sets; 4) the number of channels in the intermediate layer, and 5) the training batch size. The hyperparameters are tuned by searching for values that minimize the validation loss of the model. A minimum of five experiments were conducted for each set of hyperparameter values, with the objective of evaluating the impact of these values on the performance. 

The hyperparameters were tuned by searching for values that minimize the model’s validation loss, and the final selections are listed in Tab.~\ref{tab:hyperparameters}.
\end{enumerate}
\begin{table}[htbp]
\centering
\caption{Hyperparameter values used for the optimization.}
\label{tab:hyperparameters}
\begin{tabular}{lcc}
\hline
\hline
Hyperparameter & Minimum Value & Maximum Value  \\
Learning rate & $5\times10^{-5}$ & $3\times10^{-2}$ \\
Dropout percentage & 10 & 50  \\
Number of training datasets & 6 & 16 \\
Number of channels & 32 & 256 \\
Training batch size & 1 & 3  \\
\hline
\hline
\end{tabular}
\end{table}

\subsection{Loss function}
The objective of the training for UNet is to minimize the loss between the prediction for a given field and the simulation truth for each voxel.
\begin{enumerate}
\item[1)]
For training our UNet model to reconstruct the real-space DM density fields $\rho_{\rm DM}$, we use the Mean Squared Error (MSE) loss function,
\begin{equation}\label{eq:lossd}
\mathcal{L} = \frac{1}{N} \sum_{i=1}^N  \big(\rho_i-\rho^{\rm true}_i\big)^2 \,,
\end{equation}
where the index $i$ runs over all $N$ pixels of a field.
\item[2)]
To account for the contributions of the velocity magnitude ($v\equiv |\bm{v}|$) and the velocity direction (unit vector $\hat{v}\equiv \bm{v}/|\bm{v}|$), we choose the following two-term loss function, 
\begin{equation}
 \mathcal{L} = \frac{1}{N} \sum_{i=1}^N \left[\frac{1}{4} (|v_i|-|v_i^{\rm true}|)^2 +  \frac{3}{4}\left(1-\cos\phi_i \right) \right]\,,
 \label{eq:loss}
\end{equation}
where $\cos\phi_i\equiv  \hat{v}_i\cdot \hat{v}_i^{\rm true}$, and the index $i$ denotes the $i$-th pixel.  As observed, the first term is responsible for $|\bm{v}|$ and corresponds to the MSE loss, which is essentially equivalent to the maximum likelihood solution under the Gaussian constant variance assumption. The second term, of course, measures the deviation between the reconstructed and true values of $\hat{v}$. The coefficients of these two terms can be considered as normalization factors and are determined by the number of channels: 1 for the magnitude $|\bm{v}|$ and 3 for the three directions, i.e., $\hat{v}_x$, $\hat{v}_y$, and $\hat{v}_z$. Empirically, this loss function has proven to be stable and effective during our training process, yielding good results in velocity (momentum) reconstruction.

\begin{figure*}[htpb]
\centering
\includegraphics[width=1.9\columnwidth]{"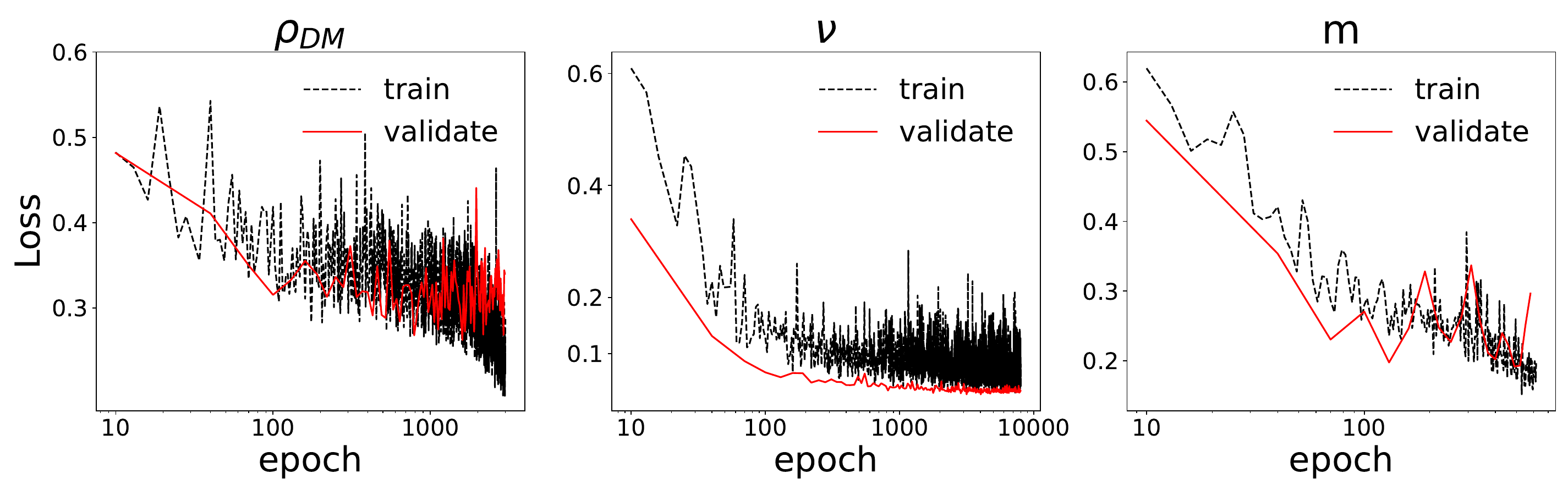"}
\caption{Loss of the training (black) and validation (red) sets. The UNnet model achieves convergence after approximately 1000 epochs of training for the DM density, DM velocity, and DM momentum (from left to right), respectively. The loss function utilized for all models is the MSE loss function, as defined in Eqs.~\ref{eq:lossd} and ~\ref{eq:loss}.}
	\label{fig:loss}
	\end{figure*}
 
\item[3)] 
We trained our UNet using the popular Adam algorithm~\citep{2014arXiv1412.6980K} for training deep neural networks. This algorithm iteratively reduces the training loss by calculating its gradient with respect to the model parameters and taking a small step in the direction of maximum reduction. From Fig.~\ref{fig:loss}, it can be seen that both the velocity model and the momentum model converge after approximately 1000 epochs of training. The main reason for the training loss fluctuations is the incorporation of dropout layers, which introduces stochastic regularization. While this helps prevent overfitting and enhances generalization, it also increases the variance in the training loss. In addition, we compared the optimal selected models and those with dropout set to zero and batch size of three for both density and the velocity field. While disabling dropout and using maximum batch size reduces training loss fluctuations, it results in noticeable overfitting and poorer validation performance. 
\end{enumerate}




\section{result}\label{sec:result}
This section will present the performance assessment of the trained UNet model, with the results presented based on predictions for 25 large boxes in the validation dataset that were not used in the model training and optimization of model structure and training parameters. Each large box have physical box size of $600^3$ $({\rm Mpc}/h)^3$ with the pixel number of $256^3$. In order to ensure the most accurate results, we selected to test on the larger boxes. This is because the measurements on the larger boxes exhibit better statistical behavior, which reduces the sampling variance and thus statistical errors.

\subsection{Visual inspection and point-wise comparison}

The main objective of this study is to compare the UNet-predicted DM velocity/momentum in real space with the ground truth. In order to facilitate the analysis of the data, it is first necessary to describe the statistics that will be employed throughout the study. The two-point correlation function is a commonly used statistical measure for characterizing a density field, 
\begin{equation}
\xi(\bm{r})=\langle\delta(\bm{x}) \delta(\bm{x}+\bm{r})\rangle\,,
\end{equation}
where the density contrast field, denoted by $\delta(\bm{x})$, is a function of any point $\bm{x}$. The separation vector, $\bm{r}$, is a vector that represents the distance between two points. The ensemble mean, represented by $\langle\cdot\rangle$, is computed by averaging over all points $\bm{x}$ in a spatial mean.
The power spectrum of $\delta(\bm{x})$ is related to the Fourier transform of the correlation function, $\xi(\bm{r})$, by a simple mathematical identity:
\begin{equation}\label{eq:pk}
P(\bm{k})=\int \xi(\bm{r}) \mathrm{e}^{i \bm{k} \cdot \bm{r}} \mathrm{d}^3 \bm{r}\,,
\end{equation}
where the three-dimensional wavevector of the plane wave, denoted by $\bm{k}$, has magnitude $k \equiv |\bm{k}|$, related to the wavelength $\lambda$ by  $k=2 \pi / \lambda$.

As with the scalar field $\delta$, we can also define power spectra for various velocity fields of interest. The velocity field, $\bm{v}$, is completely described by its divergence, $\theta \equiv \nabla \cdot \bm{v}$, and its vorticity, $\bm{\omega}=\nabla \times \bm{v}$, via the Helmholtz decomposition. In Fourier space, they become purely radial and transversal velocity modes, respectively, defined by $\theta(\bm{k})=i \bm{k} \cdot \bm{v}(\bm{k})$ and $\bm{\omega}(\bm{k})=i \bm{k} \times \bm{v}(\bm{k})$. The power spectra of the velocity, divergence, vorticity, and velocity magnitude are given by
\begin{equation}
\begin{aligned}
\langle \theta (\bm{k}) \theta^* (\bm{k}') \rangle =& (2\pi)^3 P_{\theta}(k) \delta ( \bm{k}  -  \bm{k}')\,,\\
\langle \omega^i (\bm{k}) \omega^{*j} (\bm{k}') \rangle =& (2\pi)^3 \frac{1}{2} \bigg( \delta^{ij}-\frac{k^ik^j}{k^2} \bigg) P_{\omega}(\bm{k}) \delta ( \bm{k}  -  \bm{k}')\,,\\
\langle \bm{v} (\bm{k}) \cdot \bm{v}^* (\bm{k}') \rangle =& (2\pi)^3 P_{v}(\bm{k}) \delta ( \bm{k}  -  \bm{k}')\,,
\end{aligned}\label{eq:v}
\end{equation}
where indices $i$ and $j$ denote the components in the Fourier space coordinates. 

According to the linear perturbation theory, the continuity equation leads to the following relationship: $\theta=-\mathcal{H}f\delta$, where $\mathcal{H} = aH$ is the conformal Hubble parameter, $a$ represents the cosmic scale factor, $f_g$ is the linear growth rate, defined as $f_g = d\ln D/d\ln a$, with $D$ being the linear density growth factor. Under the $\Lambda$CDM model, the growth rate is approximately given by $f_g \approx \Omega_{m}^{0.55}$~\citep{2005PhRvD..72d3529L}.

\begin{figure*}[htpb]
		\centering
		\includegraphics[width=1.9\columnwidth]{"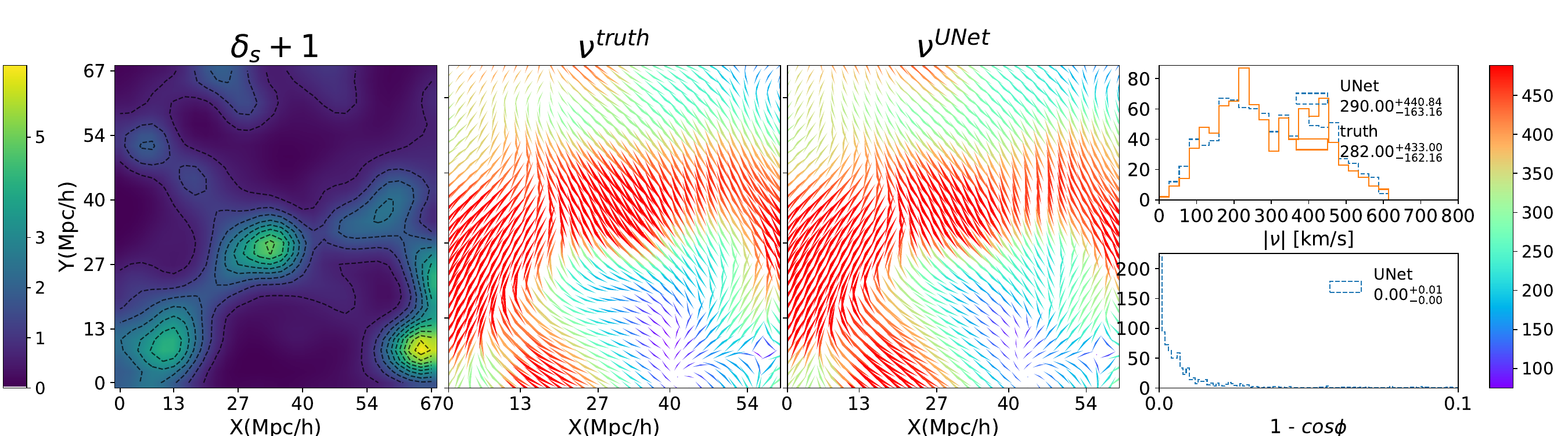"}
		\includegraphics[width=1.9\columnwidth]{"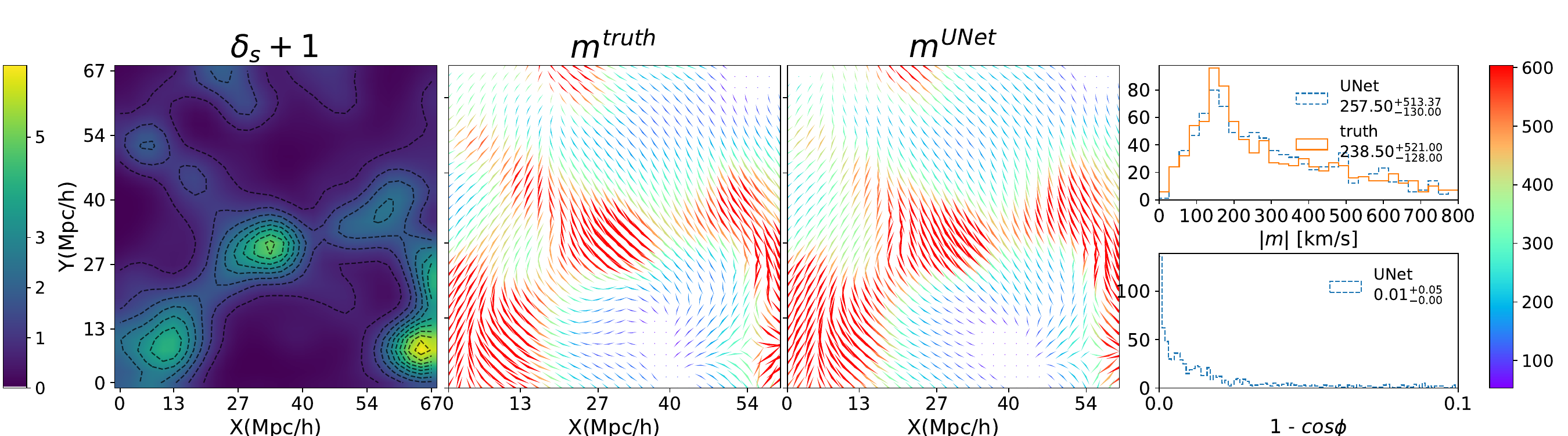"}
		\caption{Comparison of the reconstructed fields by UNet with the true fields for velocity (upper) and momentum (lower) for slices from the test set, using the ``without $M_{\rm halo}$ weighting'' weighting scheme. Each slice volume is $68\times68\times23~(h^{-1}{\rm Mpc})^3$. The sparse DM halo field  ($\delta_s$), the true velocity and the reconstructed  fields are displayed from left to right, along with the corresponding histograms. For each slice, the rightmost panel displays the statistical histogram distribution of the magnitude and direction of the velocity samples. The colored arrows in the velocity field indicate the magnitude and direction of the velocity. Through both  visual and statistical analysis of the histogram, it can be observed that even in the absence of specific mass information for each DM halo, UNet is capable of accurately reconstructing the velocity/momentum fields.}
		\label{fig:vel_mo_pixel_comparison}
	\end{figure*}

 Fig.~\ref{fig:vel_mo_pixel_comparison} presents a comparison between the velocity magnitude and direction reconstructed by UNet and the ground truth values, using the ``without $M_{\rm halo}$ weighting'' weighting scheme.  As observed, the reconstructed velocity magnitude and the true velocity magnitude are $297.22 \pm 135.31$ and $294.60 \pm 131.02$ km/s, respectively. Additionally, 
the reconstructed momentum magnitude is $315.17 \pm 209.29$ km/s, while the true one is $309.50 \pm 213.43$ km/s.  The two mean values are consistent with the true values across all slices, with significant small deviations. When considered their statistical uncertainties, these deviations are negligible.

Moreover, by defining the quantity as $\cos \phi_i \equiv \hat{v}_i \cdot \hat{v}^{\rm true}_i$, where $\hat{v}_i$ is the velocity direction reconstructed by the model and $\hat{v}^{\rm true}_i$ is the true velocity direction, one can measure the deviation between the angles. The results of the tests conducted among these slices indicate that $1 - \cos\phi_i$ is $0.01\pm 0.06$ for velocity and $0.04 \pm 0.11$ for momentum, respectively. This implies that the directions of the reconstructed velocity fields are nearly identical to those of the true velocity field, and that our model performs high reconstruction accuracy.

\begin{figure*}[htpb]
		\centering
		\includegraphics[width=1.9\columnwidth]{"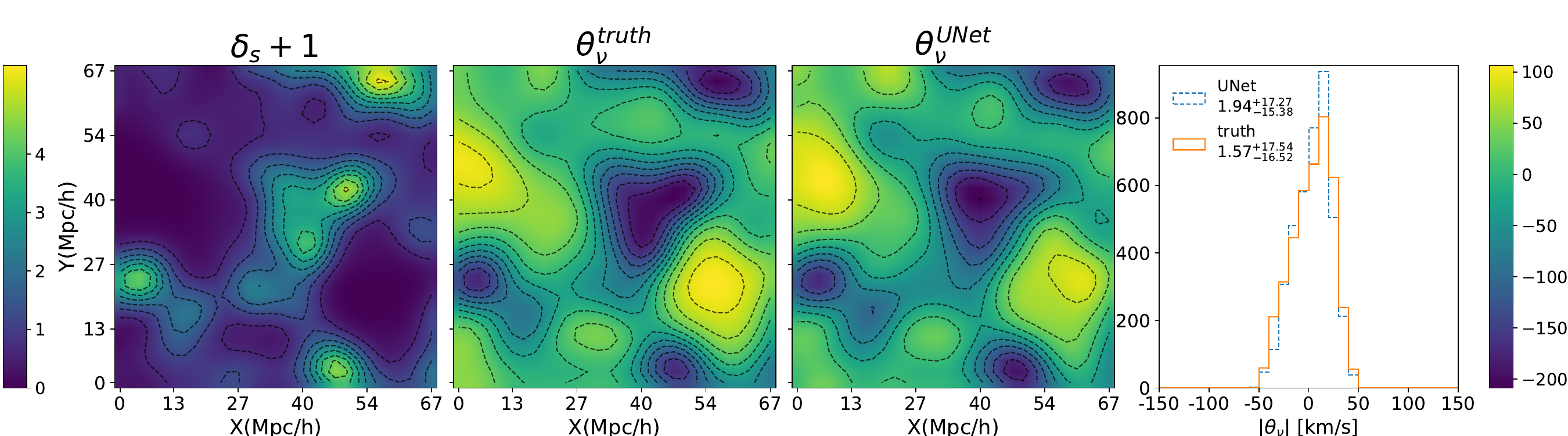"}
		\includegraphics[width=1.9\columnwidth]{"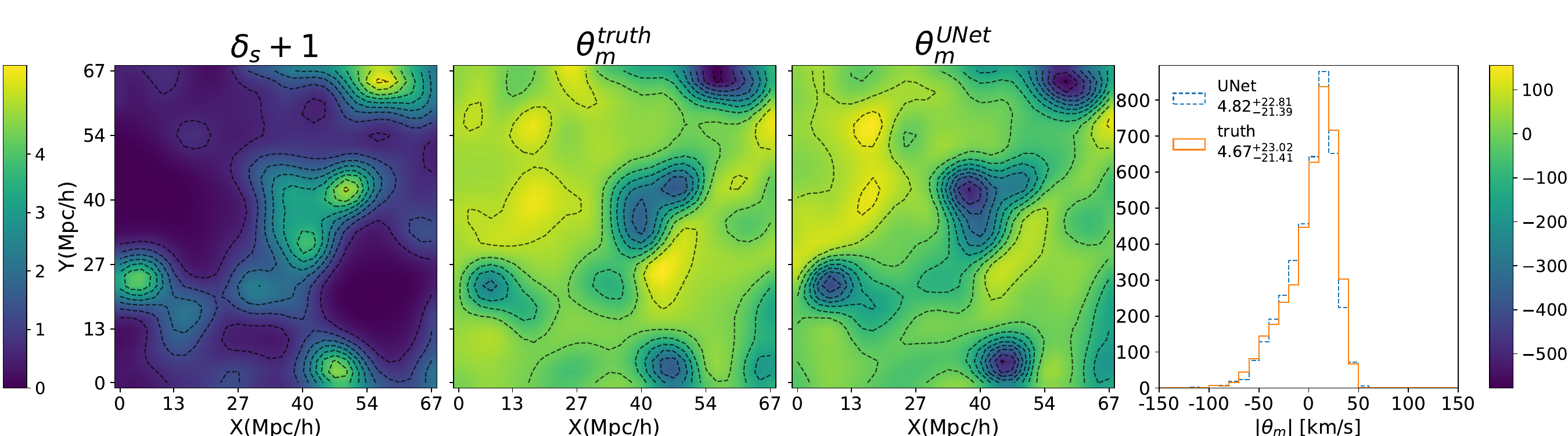"}
  \caption{Same as in Fig.~\ref{fig:vel_mo_pixel_comparison}, but for the divergence fields for velocity (upper) and momentum (lower).}
		\label{fig:vel_mo_diverge_pixel_comparison}
	\end{figure*}

Furthermore, Fig.~\ref{fig:vel_mo_diverge_pixel_comparison} illustrates a comparison between the reconstructed velocity/momentum divergence and the true velocity/momentum divergence. The divergence of velocity reconstructed by UNet and the true value are $0.40\pm 18.49$ and $0.92 \pm 17.55$ km/s, respectively. The divergence of momentum reconstructed by UNet and the true one are $0.34 \pm 25.83$ and $0.34 \pm 25.79$ km/s, respectively. At the same time, each of the histogram demonstrates that the reconstructed probability distribution is similar to that of the truth. The findings suggest that the discrepancies in the reconstruction for velocity/momentum divergence fields are well consistent with the true values, with an accuracy exceeding 1\% relative to the statistical uncertainty.

\subsection{Statistical analysis on reconstructed fields}
Furthermore, let us introduce the metrics that will be employed  for evaluating the reconstruction accuracy. For an arbitrary reconstructed field of halos from a UNet, denoted by the shorthand notation $f$, where $f \in \{\bm{v}, \theta_v\}$ for velocity and its divergence, and $f \in \{\bm{m}, \theta_m\}$ for momentum and its divergence, the following metrics are employed to describe the relative deviation and correlation coefficient, respectively, in order to compare the reconstructed field with the true one.

The relative deviation, $R$, and the correlation coefficient, $C_r$, are defined as follows:
\begin{equation}\label{eq:R}
R = \frac{\mathcal{O}_f}{\mathcal{O}_{f'}} - 1\,,\quad C_r = \frac{1}{N_{\rm pix}-1}\sum_{i}\frac{ (f_i-\bar{f}) (f'_i-\bar{f'})}{\sigma_f \sigma_{f'}}\,,
\end{equation}
where $\mathcal{O}_f$ represents an arbitrary observable for $f$. $C_r$ is defined between the reconstructed field $f$ and the true field $f'$, both of which have the same total number of pixels, $N_{\rm pix}$. The sample mean and standard deviation of field $f$ are denoted by $\bar{f}$ and $\sigma_f$, respectively. It is evident that both metrics provide a physical insight for comparison, such that the ideal reconstruction is equivalent to $R=0$ and to $C_r=1$.

\startlongtable
\begin{deluxetable*}{llcccccc}
\tablecaption{The summary of the coefficients $C_r$ and relative deviation $R$ for various components \label{tab:rc}} 
\tabletypesize{\scriptsize} 
\tablehead{ 
\colhead{Weighting Scheme} & 
\colhead{$\rho_{\rm DM}$} & 
\colhead{$|\bm{v}|$} & 
\colhead{$\theta_v$} & 
\colhead{$|\bm{m}|$} & 
\colhead{$\theta_m$} 
} 
\startdata 
$C_r$ with $M_{\rm halo}$ weighting & $0.917 \pm 0.005$ & $0.949 \pm 0.002$ & $0.948 \pm 0.001$ & $0.933 \pm 0.001$ & $0.883 \pm 0.002$ \\
$C_r$ without $M_{\rm halo}$ weighting & $0.915 \pm 0.006$ & $0.962 \pm 0.001$ & $0.947 \pm 0.001$ & $0.922 \pm 0.001$ & $0.881 \pm 0.003$ \\
\hline
$R$ with $M_{\rm halo}$ weighting & $0.018 \pm 0.001$ & $0.027 \pm 0.004$ & $-0.048 \pm 0.004$ & $-0.017 \pm 0.004$ & $-0.091 \pm 0.004$ \\
$R$ without $M_{\rm halo}$ weighting & $0.074 \pm 0.001$ & $0.033 \pm 0.004$ & $-0.03 \pm 0.004$ & $0.039 \pm 0.008$ & $-0.018 \pm 0.007$ \\
\hline
\enddata 
\tablecomments{Summary of the relative deviation, $R$, and the correlation coefficient, $C_r$, with the two weighting schemes of ``with $M_{\rm halo}$ weighting'' and ``without $M_{\rm halo}$ weighting'', for various components including the real-space DM density field ($\rho_{\rm DM}$), the magnitude of velocity, the velocity divergence, the momentum, and the momentum divergence (from left to right). For each component, the mean and the associated $1\sigma$ statistical uncertainty are shown, calculated using Eq.~\ref{eq:R}. These estimates are based on four test sets, each with a physical size of 600 ${\rm Mpc}/h$.}
\end{deluxetable*} 

The coefficients $C_r$ and relative deviation $R$ for various velocity components are calculated using Eq.~\ref{eq:R}. The calculations are based on averaging over four test sets, each with a box size of 300 ${\rm Mpc}/h$ on each side. The results are summarized in Tab.~\ref{tab:rc}. It can be observed that in all cases, the $C_r$ values for all fields, except the momentum divergence field, are at the level of 0.9. Additionally, the changes induced by the mass-weighting scheme on $C_r$ are not significant. The velocity reconstruction appears to exhibit slightly superior performance compared to the momentum reconstruction.  Moreover, the $R$ values approach zero, indicating that the reconstructed field has minimal bias compared to the true one. When comparing the ''with $M_{\rm halo}$ weighting'' scheme to the ``without $M_{\rm halo}$ weighting'' scheme, we find that even without complete knowledge of the halo mass, we can still achieve accurate reconstruction of the field.  The relatively lower $C_r$ for $\theta_m$  arises because it depends on both density and velocity, making accurate reconstruction more challenging than for either field alone. Moreover, the divergence operation further amplifies small-scale noise, further reducing the correlation.


\begin{figure*}[htpb]
	\centering
		\includegraphics[width=1\textwidth]{"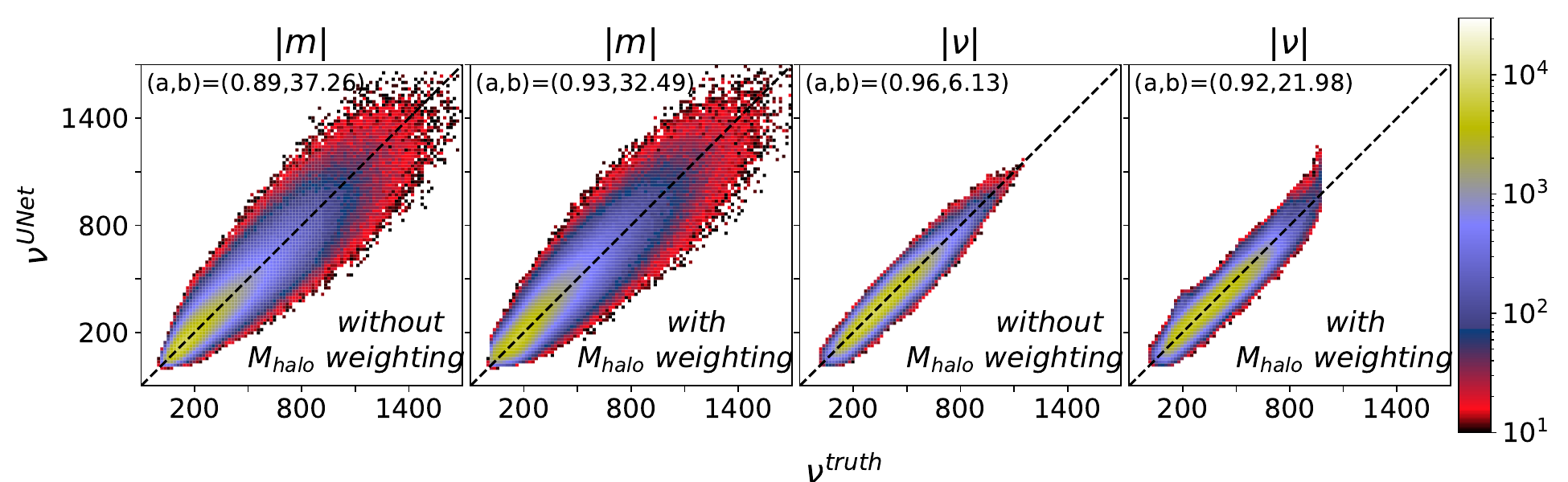"} 
	\caption{Joint probability distributions of the reconstructed fields and the true fields for velocity magnitude ($|\bm{v}|$) and momentum magnitude ($|\bm{m}|$), using the weighting schemes of ``with $M_{\rm halo}$ weighting'' and ``without $M_{\rm halo}$ weighting'' for comparison. For each field, the distributions were calculated using a large box with a side length of 600 ${\rm Mpc}/h$. The predicted distributions are strongly correlated with the true ones, indicating that the model is capable of effective reconstruction. A linear regression fit of the form $y = ax + b$, where $y$ and $x$ denote the true and reconstructed values, respectively,  is also performed for comparison. The coefficients $a$ and $b$ are presented in each plot.
 }
	\label{fig:magnitude_2joint_distribution}
	\end{figure*}

\begin{figure*}[htpb]
	\centering
		\includegraphics[width=1\textwidth]{"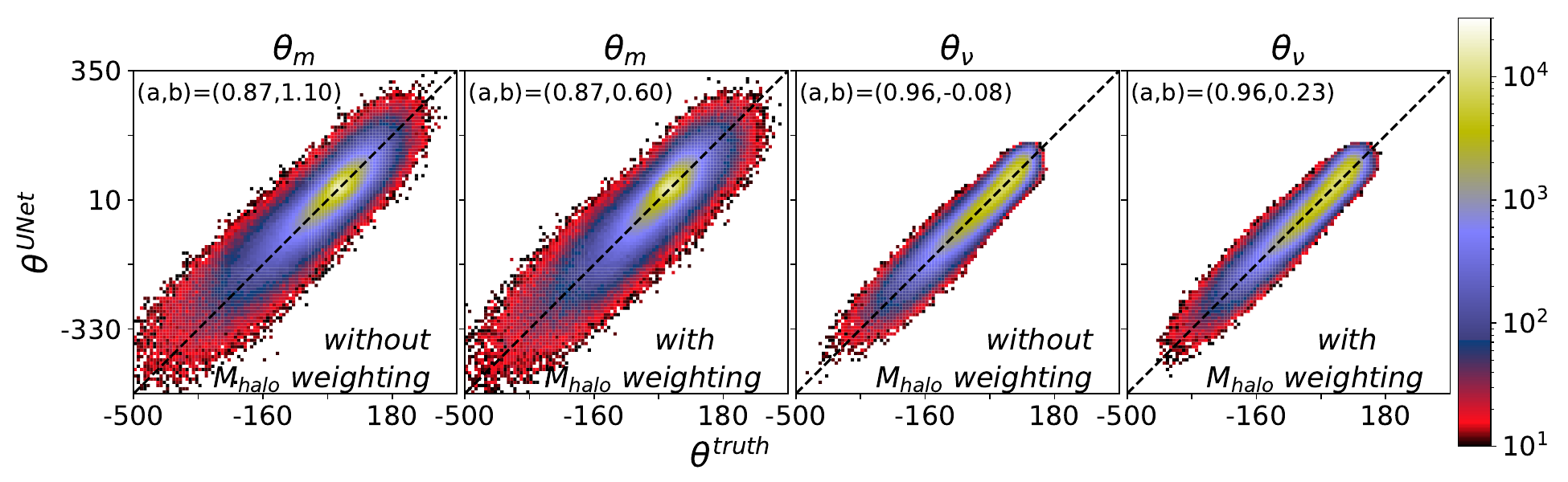"} 
	\caption{Similar to Fig~\ref{fig:magnitude_2joint_distribution}, but for the velocity divergence ($\theta_v$) and the momentum divergence ($\theta_m$), with the two mass weighting schemes. The correlation between the predicted and true distributions remains strong even when the specific mass values are not considered. This demonstrates that our model is capable of accurately predicting the distribution of DM halos even in the absence of specific mass values. These findings indicate that the mapping from the sparse DM halo number density field in redshift space to the real-space DM velocity/momentum  fields can be successfully and accurately learned by our model. }
	\label{fig:divergence_2joint_distribution}
	\end{figure*}

In order to provide a more comprehensive visual representation of the reconstruction performance, Fig.~\ref{fig:magnitude_2joint_distribution} depicts the joint probability distributions of the reconstructed fields and the true fields for magnitudes of velocity and momentum. This illustration compares the results obtained with the two mass weighting schemes. For each of the fields, the distributions were calculated using a large box with a side length of $600~{\rm Mpc}/h$. As can be observed, the predicted distributions for each field exhibit a strong correlation with the true one, falling very close to the diagonal. These findings suggest that our neural network performs an effective reconstruction. Furthermore, when employing the ``without $M_{\rm halo}$ weighting'' weighting scheme, even without knowledge of the specific halo mass values, our UNet model is still capable of accurately reconstructing the fields, closely matching the true values. This indicates that the network has learned the information about the mapping from the sparse halo number density field in redshift space to the velocity/momentum fields in real space. As demonstrated in Fig.~\ref{fig:divergence_2joint_distribution} for their divergence fields, similar findings are evident.

	\begin{figure*}[htpb]
		\centering
		\begin{tabular}{cc}
			\includegraphics[width=0.46\textwidth]{"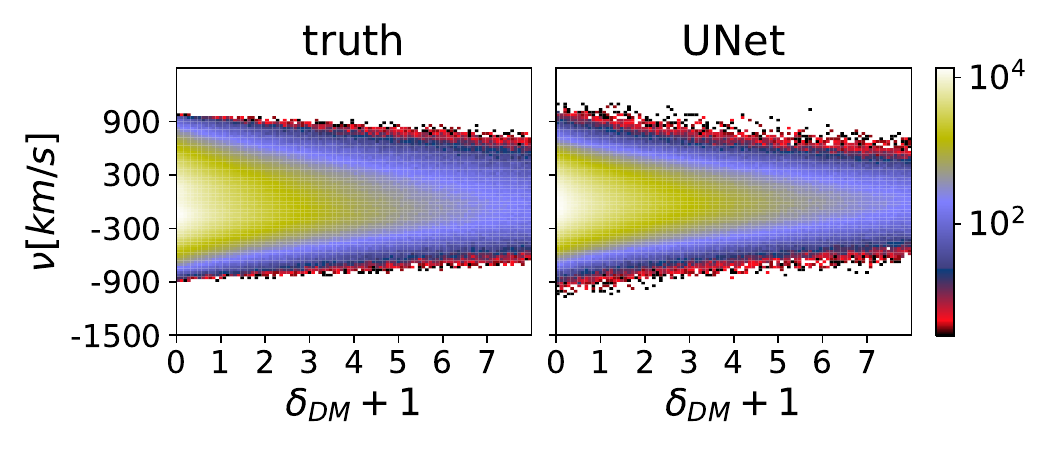"} &
			\includegraphics[width=0.46\textwidth]{"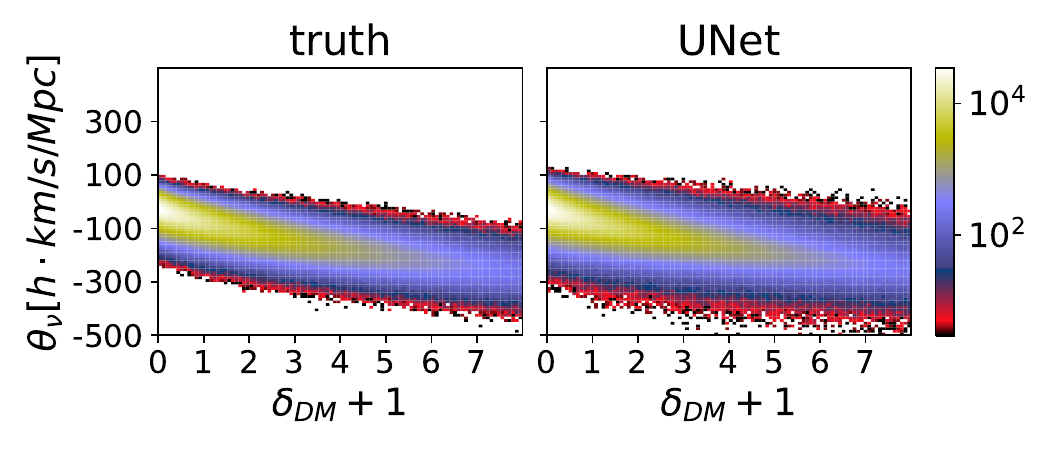"} \\
			\includegraphics[width=0.46\textwidth]{"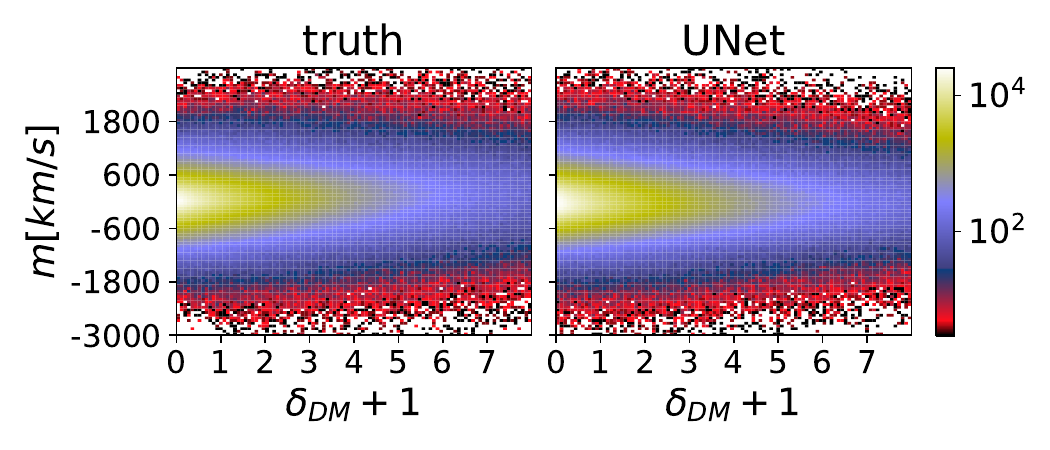"} &
			\includegraphics[width=0.46\textwidth]{"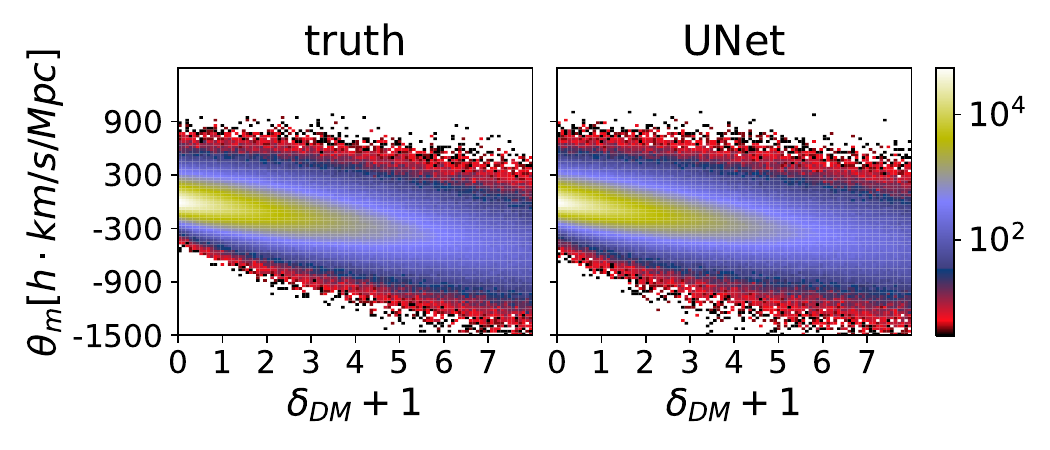"} \\
		\end{tabular}
		\caption{Comparison of the predicted joint distribution with the simulation truth for the ``with $M_{\rm halo}$ weighting'' weighting scheme. The top-left and bottom-left panels show the joint distributions of density-velocity, $\rho(\delta|v)$, and density-momentum, $\rho(\delta|m)$, respectively. The top-right and bottom-right panels show the joint distributions of density-divergence, $\rho(\delta|\theta_v)$ and $\rho(\delta|\theta_m)$, respectively. These distributions were obtained using a large box with a side length of $600~{\rm Mpc}/h$. The results demonstrate that for all DM halo number densities with $\delta + 1 \in [0, 8]$, the predicted velocity distributions exhibit good consistency with the true values.}
		\label{fig:histindelta_mass}
	\end{figure*}

	\begin{figure*}[htpb]
		\centering
		\begin{tabular}{cc}			\includegraphics[width=0.46\textwidth]{"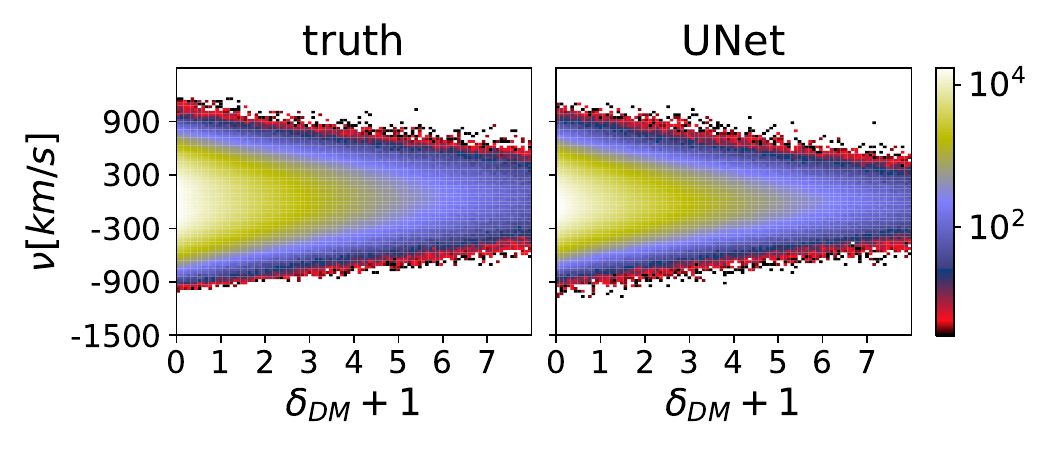"} &
		\includegraphics[width=0.46\textwidth]{"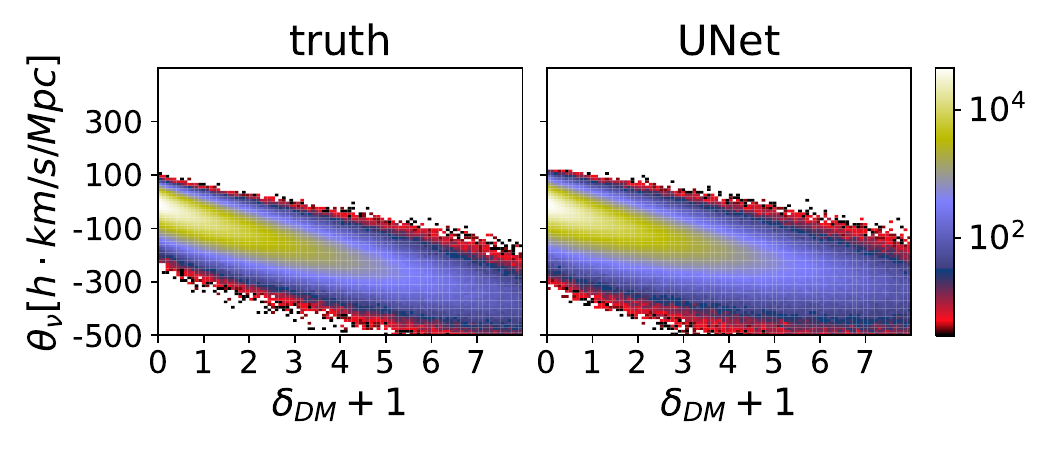"} \\		\includegraphics[width=0.46\textwidth]{"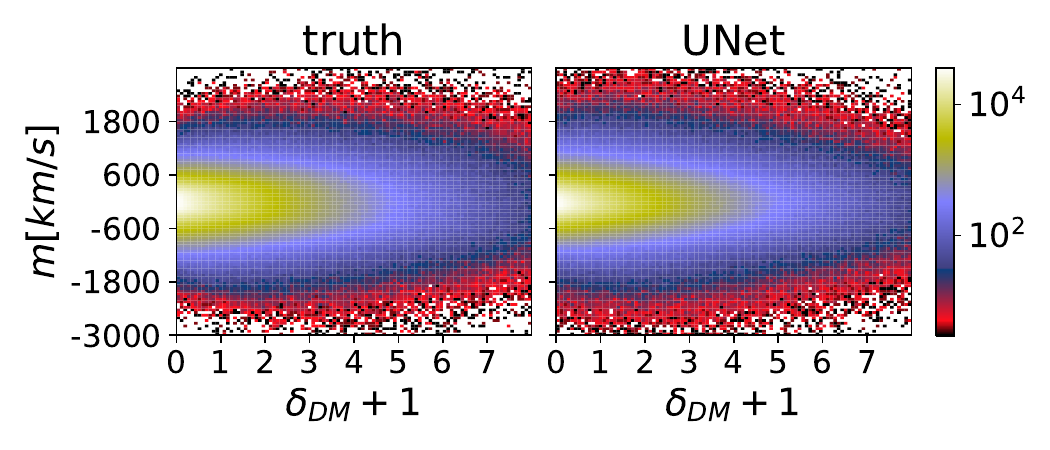"} &
		\includegraphics[width=0.46\textwidth]{"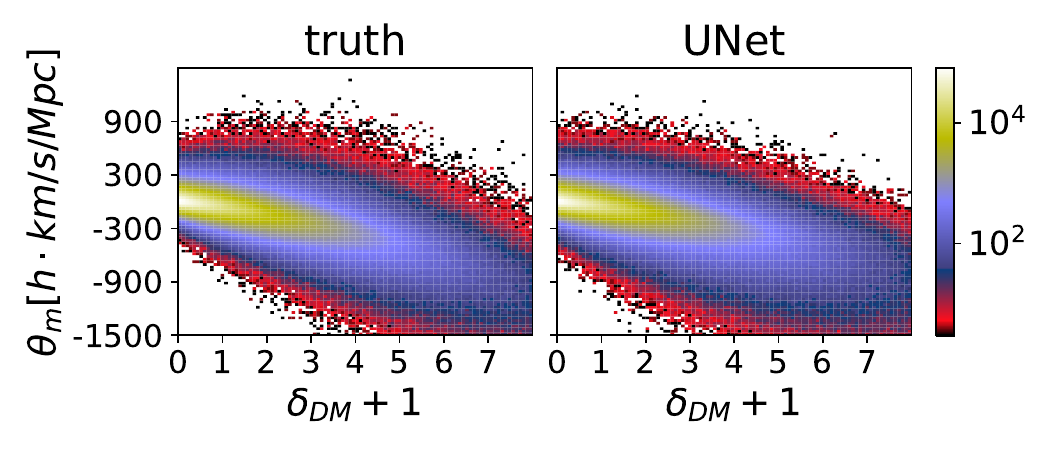"} \\
		\end{tabular}
		\caption{Same as in Fig.~\ref{fig:histindelta_mass}, but for the weighting scheme of ``without $M_{\rm halo}$ weighting''. It is evident that the predicted joint distributions remain consistent with the simulation truth, even in the absence of precise mass values.} 
		\label{fig:histindelta_nomass}
	\end{figure*}

To further test the difference between the reconstructed velocity/momentum fields and the true ones, the joint probability distribution of density-divergence and density-velocity (momentum) are shown in Fig.~\ref{fig:histindelta_mass} for the weighting scheme of ``with $M_{\rm halo}$ weighting'' and Fig.~\ref{fig:histindelta_nomass} for ``without halo mass''. It is evident that the predicted results are in strong agreement with the ground truth across a wide range of $\delta_{\rm DM}$ values, even in high-density regions where $\delta_{\rm DM} \gtrsim 4$. These high-density areas are sparse and highly non-linear, yet the model still performs effectively.

\subsection{Comparison for DM density power spectra}

 \begin{figure}[htpb]
	 	\centering
   \includegraphics[width=0.48\textwidth]{"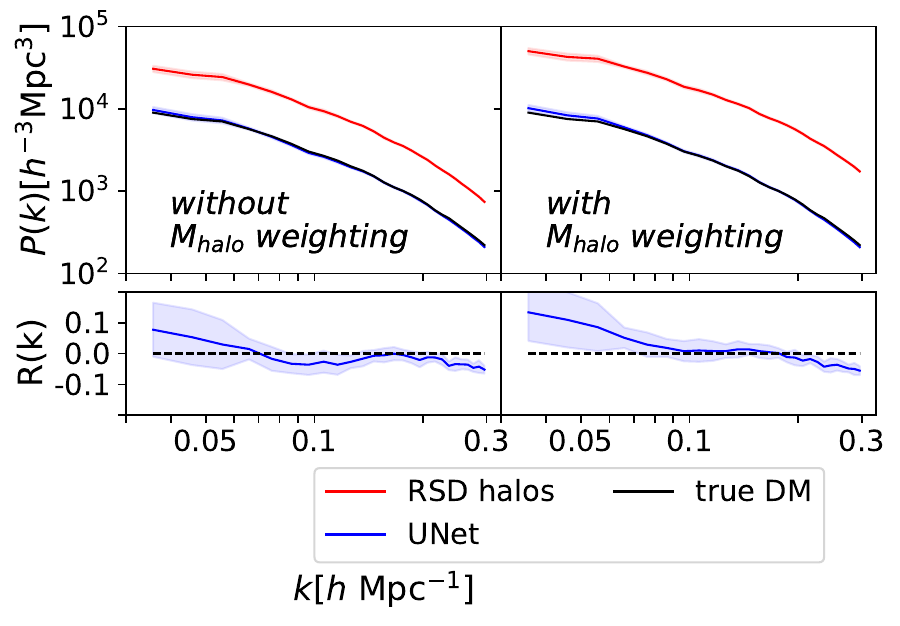"}
	 	\caption{Comparison of the UNet-predicted density power spectra with the true one. Each plot displays the predicted (blue) and true (black) values, along with the redshift-space power spectrum of the DM halos (red). The left and right panels illustrate the weighting schemes of ``with $M_{\rm halo}$ weighting'' and ``without $M_{\rm halo}$ weighting''.  A total of 25 boxes (each with a side length of $600~{\rm Mpc}/h$) were employed in the test dataset to estimate the mean and standard deviation (shaded region) for the power spectra.  The relative deviations, $R$, are presented in the bottom panels. In all cases, the UNet model yields a good reconstruction performance, insensitive to the weighting scheme. Thus, using DM halo mass as a weight does not significantly affect the reconstruction of the DM density field.}
	 	\label{fig:ps_dmp}
	 \end{figure}

As illustrated in Fig.~\ref{fig:ps_dmp}, the resulting DM density power spectra in real space, calculated using Eq.~\ref{eq:pk} for various cases, are summarized. Two mass weighting schemes of ``with $M_{\rm halo}$ weighting'' and ``without $M_{\rm halo}$ weighting'' are presented in the left and right panels, respectively, for comparison.  The auto power spectra for the predicted density (blue) and the true density (black) are shown, along with the auto power spectrum of sparse DM halo (red), providing a detailed comparison.

To accurately estimate the statistical uncertainty $\Delta P(k)$, the number of independent Fourier modes is taken into account, which leads to a rescaling of the standard deviation $\delta P(k)$ (estimated over the 25 test boxes) via
\begin{equation}\label{eq:var}
\Delta P(k)=\delta P(k) \sqrt{\frac{V_{\rm all}}{V_{\rm overlap}}}\,.
\end{equation}
Here, $V_{\rm all}$ represents the total independent volume, while $V_{\rm overlap}$ denotes the overlap volume for the calculated power spectrum. Since the  boxe have a physical size of $1200 \times 1200 \times 600~({\rm Mpc}/h)^3$, divided into 25 test boxes with a size of 600 ${\rm Mpc}/h$ on each side, the value of $\sqrt{V_{\rm all}/V_{\rm overlap}}$ is  0.4.

The relative deviation $R$ as defined in Eq.~\ref{eq:R} are illustrated in the bottom panels. For ``without $M_{\rm halo}$ weighting'' case, the discrepancies between the measured auto spectra and the true auto spectrum are significant small, essentially invisible in the left panel when $k<0.3~h/{\rm Mpc}$. 
It can be observed that the discrepancy across this scale range is $|R| < 0.13$, indicating a good recovery of both the amplitude information of the field, consistent with the findings in~\cite{Wang:2023hgm}.


In comparison to the left panels, the reconstructed power spectra for ``with $M_{\rm halo}$ weighting'' exhibit relatively larger deviations. The maximum deviation is $R=0.15$ for auto correlation. By comparing these results, we demonstrate that the absence of this information does not significantly reduce reconstruction accuracy. 

\subsection{Comparison for velocity/momentum power spectra}

	\begin{figure*}[htpb]
		\centering
	\begin{tabular}{cc}
        \includegraphics[width=0.48\textwidth]{"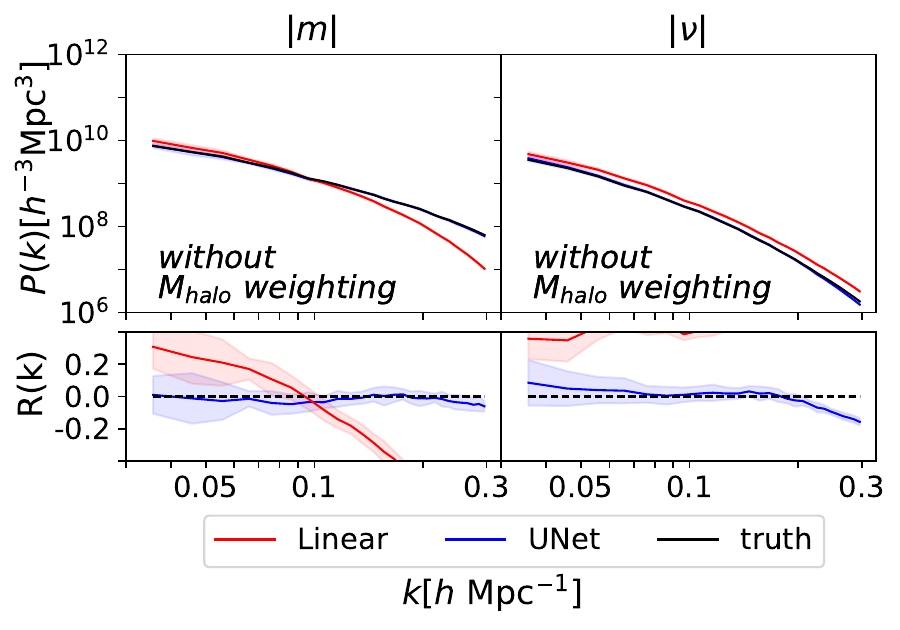"} 
        \includegraphics[width=0.48\textwidth]{"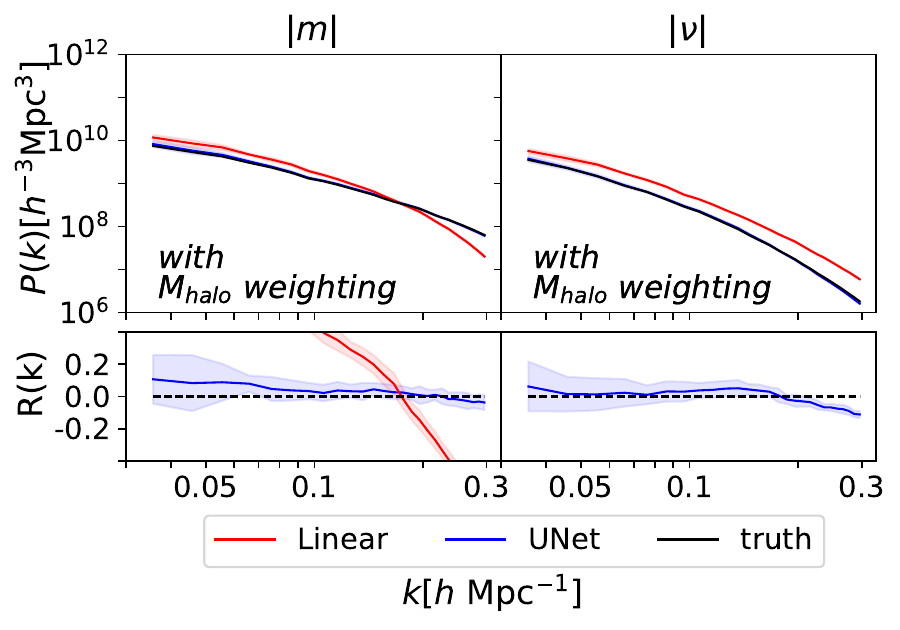"} 
        \end{tabular}
	 \caption{ Comparison of the UNet-reconstructed momentum/velocity power spectra with the simulation truth for the two mass weighting schemes. The specific field and the weighting scheme are clearly indicated in each panel. The bottom panels present the resulting relative deviations. In each plot, the shaded region represents the $2\sigma$ statistical uncertainty, estimated from a total of 25 boxes (each with a side length of $600~{\rm Mpc}/h$) in the test dataset. Based on Eqs.~\ref{eq:linear} and~\ref{eq:linear_momentum},  the linear predictions are also provided for comparison, deviating increasingly from the true values as 
$k$ increases.  As seen, the relative deviations are in the range of $R \in [-0.06,0.06]$ at the scales of $k \in [0.05,0.1]$. The average relative deviations of auto power spectrum are $-0.013 \pm 0.03, -0.011 \pm 0.07, 0.033 \pm 0.05, -0.004 \pm 0.053$ from left to right, respectively. }
	 \label{fig:ps_vel_amplitude}
 	\end{figure*}

 \begin{figure*}
 	\centering
	\begin{tabular}{cc}
        \includegraphics[width=0.48\textwidth]{"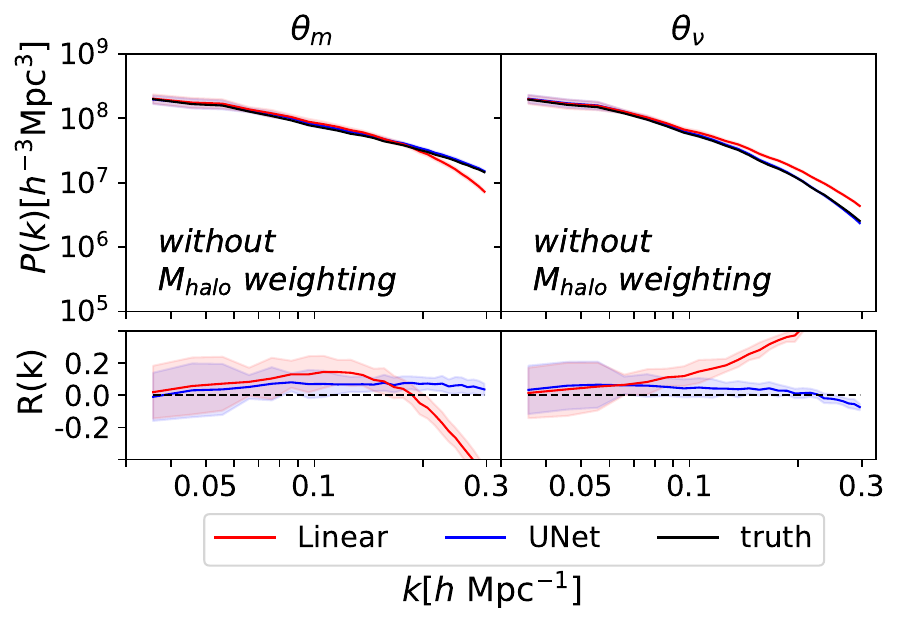"} &
        \includegraphics[width=0.48\textwidth]{"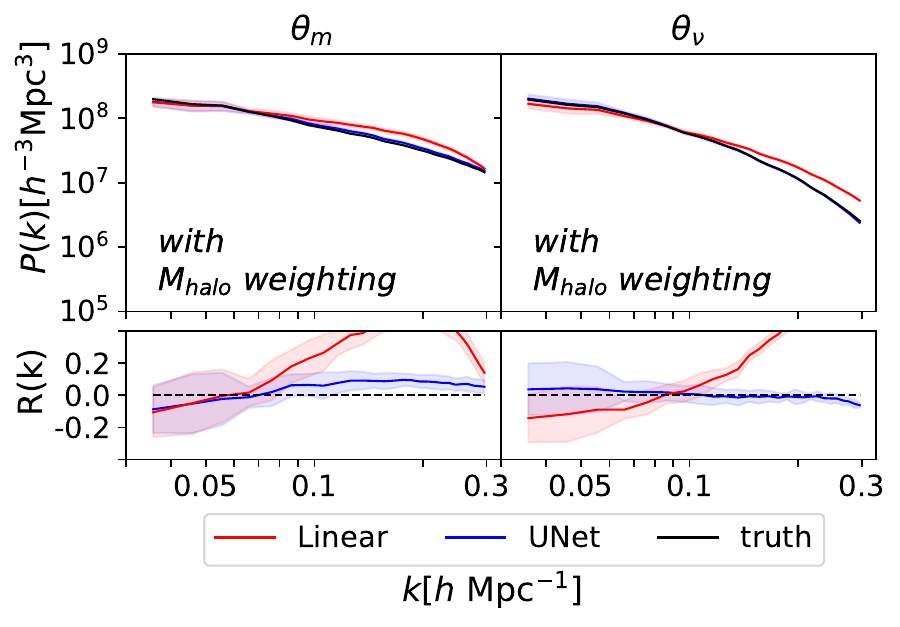"} 
        \end{tabular}
 	\caption{Same as in Fig.~\ref{fig:ps_vel_amplitude}, but for the momentum/velocity divergence fields with two different mass weighting schemes. The specific field and the weighting scheme are clearly indicated in each panel. The relative deviations fall within the range of $[-0.1, 0.1]$ at the scale of $k \in [0.05 , 0.3]$, respectively, indicating a relatively accurate reconstruction. Compared to the linear predicted power spectra, the UNet model provides a more accurate reconstruction, especially when $k>0.1~h/{\rm Mpc}$. }
 	\label{fig:ps_vel_divergence}
 \end{figure*}

Furthermore, we now examine the power spectrum reconstruction for velocity, momentum, and their divergences. In Fig.~\ref{fig:ps_vel_amplitude}, the magnitudes of the momentum and velocity power spectra are presented. The auto correlation power spectra between the predicted and true fields based on the two weighting schemes are also shown for comparison. The estimated curves and shaded regions are based on the mean and standard deviation of the derived power spectra from a total of 25 boxes in the test dataset, each with a side length of $600~{\rm Mpc}/h$.      

It can be observed that the auto correlation power spectra of the velocity and momentum can be reconstructed with a relative deviation of $|R(k)|<0.07$ for $k \in [0.05,0.2]~h{\rm Mpc}^{-1}$. In contrast, the velocity and momentum spectra have $|R(k)|\in[0,0.13]$, indicating an overestimate in the scales of $k\in[0.03,0.05]~h{\rm Mpc}^{-1}$.  In general, the reconstruction performance for ``with $M_{\rm halo}$ weighting'' across all scales appears better than that for ``without $M_{\rm halo}$ weighting''. This is evidenced by the fact that even without complete knowledge of the halo mass, we can still accurately reconstruct the auto-correlation of the DM velocity field and momentum field using UNet.

Moreover, since the momentum field is a density-weighted velocity, it is expected to be more sensitive to the reconstruction accuracy of the DM density field. However, the reconstruction accuracy for the momentum is comparable to that of the velocity, thanks to the good reconstruction of the DM density, as illustrated in Fig.~\ref{fig:ps_dmp}. It is clear that as the reconstruction uncertainty from the DM density field increases, the statistical error in the momentum and its divergence power spectra will also increase. 

Moreover, as shown in Fig.~\ref{fig:ps_vel_divergence}, the auto power spectra of the velocity and momentum divergence yield $R(k) < 0.06$ for scales of $k$ from $0.05$ to 0.1 $h/{\rm Mpc}$. At $k > 0.05$ $h/{\rm Mpc}$, the auto-correlation of velocity divergence agrees well with the true power spectrum within $2\sigma$, while the auto-correlation of momentum divergence slightly overestimates the true value.  We speculate that the excess in the momentum divergence power spectrum (for both ``with/without halo mass'' schemes) arises from the divergence operation amplifying noise across scales during UNet learning, adding a positive contribution to the auto power spectrum.


In addition, a comparison of the results obtained by the two different weighting schemes reveals that there is not a significant difference between them. This suggests that precise mass information is not necessary for the velocity/momentum reconstruction. Overall, our UNet model demonstrates satisfactory performance on linear and nonlinear scales of $k\in [0.03,0.3]~h/{\rm Mpc}$. These results highlight the ability of the UNet model to learn various velocity quantities from the halo number density field on the nonlinear scales, which is an important ﬁnding in this study. In light of the predicted power spectra, it can be reasonably concluded that the UNet model's predictions are in relatively good agreement with the truth. 

\subsection{RSD corrections and power spectrum multipoles}
To further validate the effectiveness of the UNet model, we will demonstrate comparisons between the reconstructed measurements and the simulation truth in real space. As the input observables in the neural network are in redshift space, the relevant comparison can be used to assess the accuracy of the peculiar velocities corrected. The following analysis was conducted based on a total of 18 simulation boxes, each with a side length of $600~{\rm Mpc}/h$.

\begin{figure}[!h]
\centering
\includegraphics[width=0.45\textwidth]{"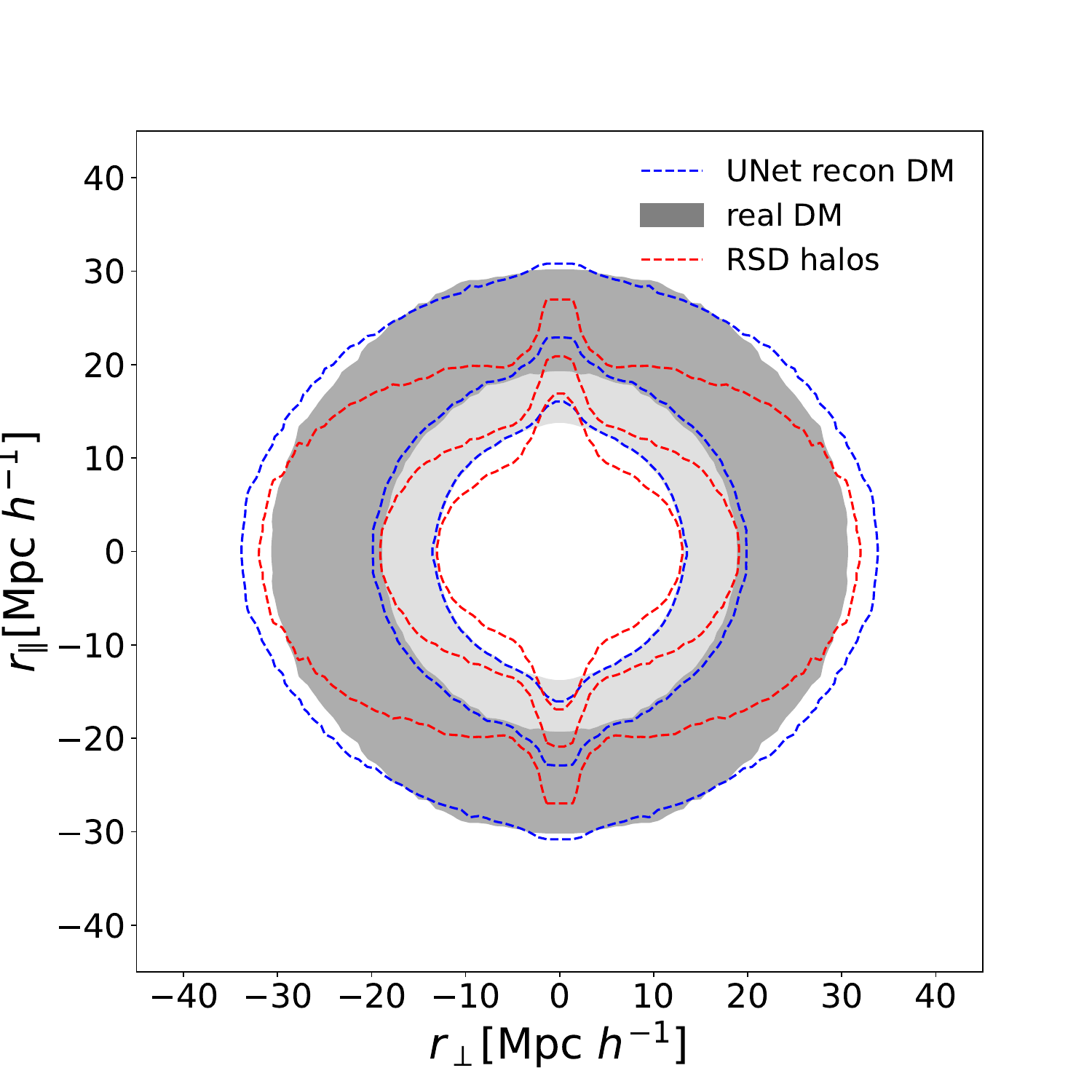"} 
\caption{Comparison of the two-dimensional anisotropic 2PCFs, $\xi(r_{\perp}, r_{\|})$, of density fields between the prediction and the simulation truth in real space for ``with $M_{\rm halos}$ weighting'' scheme. The results are based on a total of 25 boxes, each with a side length of $600~{\rm Mpc}/h$. The contours were generated from the correlation values of $[0.02,0.06,0.12]$, respectively. The gray shaded regions and blue dashed contours represent the derived $\xi$ of the true and reconstructed fields in real space, respectively. The red dotted contours represent the redshift-space 2PCF of the DM halos. It is evident that despite the input being that of the DM halo density in redshift space, the reconstructed 2PCF is in good agreement with the true value. This suggests that the UNet is capable of successfully recovering the real-space DM density field.}
\label{fig:2d2pcf}
\end{figure}

In Fig.~\ref{fig:2d2pcf}, a comparison is performed between the two-dimensional anisotropic 2PCFs, $\xi(r_{\perp}, r_{\|})$, of density fields between the prediction and the simulation truth in real space. As illustrated by the red dotted contours, the 2PCF measured in redshift space without any RSD corrections demonstrates the significant impact of RSD effects. The Kaiser effect results in the ``squashed'' structure along LoS, while the random velocities on the non-linear small scales produce the so-called ``fingers-of-God'' (FoG) effect, which causes structures to be elongated along the LoS.  The velocity field reconstructed by UNet was utilized to correct the velocities of the halo. Note that the corrected 2PCF exhibits a high degree of agreement with the real-space 2PCF across all scales. Furthermore, the 2PCF corrections for both Kaiser and FoG effects are achieved with a high degree of accuracy. The successful corrections for the FoG effect indicate that the UNet is capable of accurately predicting the DM field in real space.   

In order to accurately assess the statistical properties of an arbitrary real-space field, it is necessary to consider the higher-order power spectrum multipoles. In general, the power spectrum in redshift space $P(k, \mu)$, where $\mu$ is the cosine of the angle between the wavevector $\bm{k}$ and the LoS direction, can be expressed by expanding its multipole components:
\begin{equation}\label{eq:multipoles}
P_\ell(k)=\frac{2 \ell+1}{2} \int_{-1}^1 d \mu P(k, \mu) \mathcal{L}_\ell(\mu)\,,
\end{equation}
where $\mathcal{L}_\ell$ are the Legendre polynomials of order $\ell$. In this study, we focus on the Legendre polynomials of order $\ell$, specifically the quadrupole ($\ell=2$) and the hexadecapole ($\ell=4$), denoted as $P_{2}$ and $P_{4}$, respectively.

\begin{figure*}[htpb]
    \centering
	\begin{tabular}{cc}
    \includegraphics[width=0.48\textwidth]{"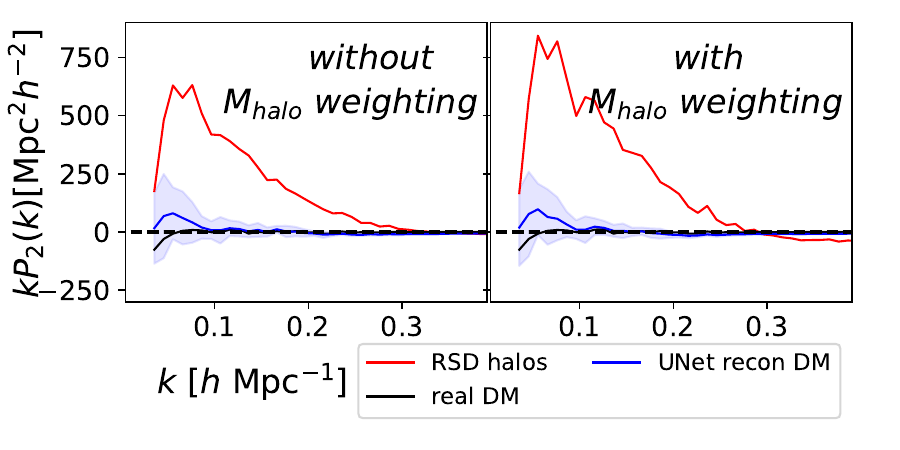"}&
    \includegraphics[width=0.48\textwidth]{"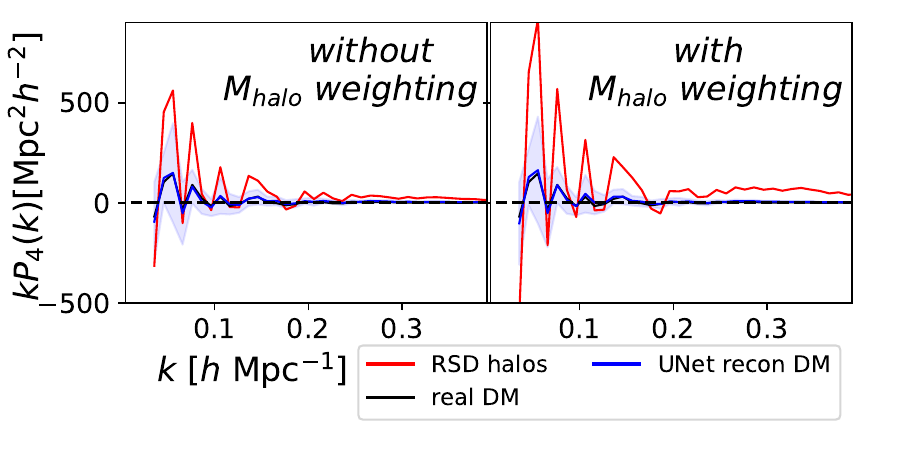"} 
        \end{tabular}
    \caption{Comparison of the UNet-reconstructed power spectrum multipoles of the DM fields with the truth is presented. Top: a comparison of the measured quadrupole ($P_2$) of the DM halos in redshift space (red dashed), the UNet-reconstructed DM field (blue dashed) and the true value (black solid) in real space. The two mass weighting schemes: ``without $M_{\rm halo}$ weighting'' (left) and ``with $M_{\rm halo}$ weighting'' (right) are chosen to illustrate the reconstructed performance. The shaded regions represent the $2\sigma$ confidence level, estimated from the test dataset. Bottom: same as in the top panels, but for the higher-order multipole of hexadecapole ($P_4$).  As demonstrated, following the application of RSD corrections, the reconstructed multipoles of the quadrupole and hexadecapole exhibit well agreement with the true values at the $2\sigma$ level within the range of $k\in[0.04,0.4]~h/{\rm Mpc}$. Furthermore, the scheme of ``with $M_{\rm halo}$ weighting'' is observed to provide a higher reconstruction accuracy and a smaller statistical uncertainty.}
    \label{fig:dmp_ps_multipole}
\end{figure*}

In Fig.~\ref{fig:dmp_ps_multipole}, we compare the power spectrum multipoles of the UNet-reconstructed DM density field with the true values. The top panels depict the measured $P_2$ in redshift space (red), the UNet-reconstructed DM field (blue) in real space, and the real-space true value (black). Two mass weighting schemes were employed for the purpose of demonstrating the reconstruction performance: ``without $M_{\rm halo}$ weighting'' (left) and ``with $M_{\rm halo}$ weighting'' (right). The shaded regions represent the $1\sigma$ confidence interval, estimated based on the test dataset. 

 As demonstrated, the RSD effects can markedly contribute to anisotropic clustering, resulting in an amplitude of $P_2 (k)$ that is approximately one to two orders of magnitude greater than that observed in real space across all $k$ values of interest. The quadrupole of the UNet-predicted DM density field in real space can significantly suppress the anisotropy clustering, yielding a reconstructed result that is in good agreement with the true one at the $2\sigma$ level when $k \lesssim 0.4~h/{\rm Mpc}$. Furthermore, it has been found that the ``with $M_{\rm halo}$ weighting'' scheme yields a slightly more accurate reconstruction. The discrepancy between the reconstructed and true values averaged over the $k$ range is $174.29 \pm 702.27$ for the ``with $M_{\rm halo}$ weighting'' scheme and $175.05 \pm 711.88$ for the ``without $M_{\rm halo}$ weighting'' scheme. This indicates that we can reconstruct the quadrupole of the DM density field without complete knowledge of halo mass information.  

The right panels of Fig.~\ref{fig:dmp_ps_multipole}  depict a similar comparison of a higher-order multipole, the hexadecapole, $P_4$. As a consequence of its higher order, the statistical uncertainty for halos in redshift space appears to exceed that of $P_2$. Nevertheless, the UNet is still capable of successfully reconstructing  $P_4$ of the DM density field in real space, with the associated statistical uncertainty being approximately comparable to that of $P_2$. The discrepancy between the reconstructed and true values falls within the $2\sigma$ level. Furthermore, the use of a weighting scheme based on halo mass can result in a slight improvement in reconstruction accuracy, with the average deviation of $-60.03 \pm 543.83$ reduced by $27\%$ compared to the ``without $M_{\rm halo}$ weighting'' case.

The UNet-based RSD correction is less effective in the small $r_\perp$ region, where the FoG effect dominates. To quantify its impact on the reconstructed quadrupole and hexadecapole moments, we applied a Gaussian smoothing filter along LoS with a standard deviation of $\sigma_z = 3.5~{\rm Mpc}/h$, effectively suppressing small-scale FoG distortions. This filtering significantly improves the isotropic 2PCF reconstruction at $r_\perp < 10~{\rm Mpc}/h$. In Fourier space, the difference between the multipole spectra with and without smoothing remains small; for example, at $k = 0.1~h/{\rm Mpc}$, the relative deviation between the unsmoothed and smoothed power spectra, in units of the associated standard deviation, is around 0.1 across all relevant scales. This indicates that the FoG effect introduces only minor residuals in the multipole power spectra, and while UNet cannot directly correct for it, the overall impact is negligible.

The proposed UNet is capable of automatically applying RSD corrections. Furthermore, the reconstructed multipoles of both the quadrupole and hexadecapole have been found to agree with the true values at the $2\sigma$ level, within the range of $k \in [0.06, 0.3]~h/{\rm Mpc}$. Moreover, the ``with $M_{\rm halo}$ weighting'' scheme has been observed to yield higher reconstruction accuracy and smaller statistical uncertainty.

In conclusion, while the RSD effects can significantly contribute to anisotropic clustering, distorting $P_2(k)$ and $P_4(k)$, our proposed UNet effectively corrects for these RSD effects and yields accurate predictions for the real-space statistical measurements of density. The reconstructed results align well with the true values at the $2\sigma$ level for $k \in [0.03, 0.4]~h/{\rm Mpc}$. Furthermore, the ``with $M_{\rm halo}$ weighting'' scheme provides additional information, resulting in a slightly more accurate reconstruction.

\section{Conclusions}\label{sec:con}
The construction of three-dimensional velocity (and momentum) fields by galaxies and clusters is of great significance in the field of cosmology. These fields provide a wealth of information that is not available from the density field alone. They can be used to improve and correct various cosmological measurements. The reconstruction of high fidelity may even result in the discovery of unexpected findings. The reconstruction of cosmic volume-weighted velocity is susceptible to significant sampling artifacts, which presents a challenge for traditional reconstruction methods. Furthermore, traditional reconstruction methods frequently rely on numerous assumptions and approximations.   

The objective of this study is to propose a deep learning approach based on the UNet neural network for the reconstruction of three-dimensional velocity/momentum fields in real space. This approach may provide a potential solution to the long-standing problem. This study has demonstrated the effectiveness of UNet in reconstructing fields directly from sparse samples of DM halo (and subhalo) spatial distribution in redshift space. The UNet architecture, with its sophisticated design, is capable of capturing diverse field characteristics and transforming high-dimensional, structured inputs, making it a valuable tool for such reconstructions. 


The UNet was trained to transform the sparse halo density fields into velocity/momentum fields. To include the halo mass information and assess the sensitivity of the reconstruction performance to this knowledge, we adopted two weighting schemes: one that includes halo mass and one that does not. A comprehensive validation was conducted through a series of statistical tests, and the reconstructed velocity/momentum fields were found to exhibit a high degree of agreement with the ground truth. Moreover, the UNet-inferred DM velocity fields in real space essentially provide effective RSD corrections. Compared to the true values, we have demonstrated that the proposed UNet can reconstruct real-space velocity and momentum fields with a relative deviation of about $R < 0.13$ in highly non-linear regimes at $k < 0.3~h/{\rm Mpc}$.

It is also worth noting that the reconstruction of our UNet model is effective even in the absence of precise DM halo mass information. A comparison of the pixel-by-pixel values and power spectra reveals that the UNet model is still capable of accurately reconstructing not only the DM density field and relevant power spectrum, but also the velocity/momentum fields and the corresponding power spectra. The reconstruction accuracy is not significantly different from those obtained when the accurate DM halo mass value is used. The findings indicate that the model is capable of performing reconstruction in a manner that is more consistent with real observations, which is a highly encouraging indication for the reconstruction of velocity/momentum fields using current and future real data.

The Stage IV galaxy surveys will yield more detailed measurements of LSS of the Universe than ever before. In consequence, novel computing technologies are required to analyze these high-dimensional, massive data sets. It is therefore anticipated that UNet-based neural networks will prove an invaluable tool in addressing the challenges inherent in traditional methodologies, thereby facilitating the extraction of cosmological information in a more profound and comprehensive manner.

\section*{Acknowledgments}
This work is supported by National SKA Program of China (2020SKA0110401, 2020SKA0110402, 2020SKA0110100), the National Key R\&D Program of China (2020YFC2201600), the National Science Foundation of China (12203107, 12073088, 12373005, 12103037 ), the China Manned Space Project with No. CMS-CSST-2021 (A02, A03, B01), the Guangdong Basic and Applied Basic Research Foundation (2024A1515012309),the Fundamental Research Funds for the Central Universities,Sun Yat-Sen University(No.24qnpy122) and the 111 project of the Ministry of Education No. B20019. We also wish to acknowledge the Beijing Super Cloud Center (BSCC) and Beijing Beilong Super Cloud Computing Co., Ltd (\url{http://www.blsc.cn/}) for providing HPC resources that have significantly contributed to the research results presented in this paper.

\bibliography{sample631}{}
\bibliographystyle{aasjournal}



\end{document}